\documentclass[namedreferences]{solarphysics}
\usepackage[optionalrh]{spr-sola-addons} 
\usepackage{graphicx}        
\usepackage{color}           
\usepackage{breakurl}        
\usepackage{sidecap}

\renewcommand{\vec}[1]{{\mathbfit #1}}

\newcommand{\curl}{ {\bf \nabla} \times}

\chardef\us=`\_

\newcommand{\fig}[1]{Figure~\ref{fig:#1}}

 \newcommand{\sect}[1]{Sect.~\ref{sec:#1}}
 \newcommand{\sects}[2]{Sects.~\ref{sec:#1} and \ref{sec:#2}}

\newcommand{\arcsec}{''} 
\newcommand{\degree}{^{\circ}} 
\newcommand{\Rsun}{$R_{\odot}$}

\usepackage[bookmarks = true,colorlinks = true,linkcolor = blue,urlcolor = blue,citecolor = blue,bookmarks = true,linktocpage = true,hyperindex = true,breaklinks]{hyperref}

\usepackage[dvipsnames]{xcolor}
\definecolor{Mycolor2}{HTML}{00F9DE}

\DeclareUnicodeCharacter{2212}{-}

\begin{document}

\begin{article}
\begin{opening}

\title{
Analysis of the Evolution of a Multi-Ribbon Flare and Failed Filament Eruption}

\author[addressref={aff1,aff2,aff3},corref,email={reetika.joshi@astro.uio.no}]{\inits{R.}\fnm{Reetika}~\lnm{Joshi}\orcid{0000-0003-0020-5754}}
\author[addressref={aff4},corref,email={mandrini@iafe.uba.ar}]{\inits{C.H.}\fnm{Cristina H.}~\lnm{Mandrini}\orcid{0000-0001-9311-678X}}
\author[addressref={aff3},corref,email={rchandra.ntl@gmail.com}
]{\inits{R.}\fnm{Ramesh}~\lnm{Chandra}\orcid{0000-0002-3518-5856}}
\author[addressref={aff5,aff6,aff7},corref,email={Brigitte.Schmieder@obspm.fr}
]{\inits{B.}\fnm{Brigitte}~\lnm{Schmieder}\orcid{0000-0003-3364-9183}}
\author[addressref={aff4},corref,email={gcristiani@gmail.com}]{\inits{G.}\fnm{Germ\'an D.}~\lnm{Cristiani}\orcid{0000-0003-1948-1548}}
\author[addressref={aff4},corref,email={ceciliamaccormack@gmail.com}]{\inits{C.}\fnm{Cecilia}~\lnm{Mac Cormack}\orcid{0000-0003-1173-503X}}
\author[addressref={aff5,aff8},corref,email={Pascal.Demoulin@obspm.fr}]{\inits{P.}\fnm{Pascal}~\lnm{D\'emoulin}\orcid{0000-0001-8215-6532}}
\author[addressref={aff9,aff10},corref,email={hebe.cremades@gmail.com}]{\inits{H.}\fnm{Hebe}~\lnm{Cremades}\orcid{0000-0001-7080-26642}}


\address[id=aff1]{Institute of Theoretical Astrophysics, University of Oslo, P.O. Box 1029 Blindern, N-0315 Oslo, Norway}
\address[id=aff2]{Rosseland Centre for Solar Physics, University of Oslo, P.O. Box 1029 Blindern, N-0315 Oslo, Norway}
\address[id=aff3]{Department of Physics, DSB Campus, Kumaun University, Nainital 263-001, India}
\address[id=aff4]{Instituto de Astronomía y Fïsica del Espacio, IAFE, UBA-CONICET, CC.67, Suc.28,1428 Buenos Aires, Argentina}
\address[id=aff5]{LESIA, Observatoire de Paris, Université PSL, CNRS, Sorbonne Université, Université de Paris, 5 place Jules Janssen, 92190 Meudon, France}
\address[id=aff6]{Centre for Mathematical Plasma Astrophysics, Dept. of Mathematics, KU Leuven, 3001 Leuven, Belgium}
\address[id=aff7]{SUPA, School of Physics \& Astronomy, University of Glasgow, G12 8QQ, UK}
\address[id=aff8]{Laboratoire Cogitamus, rue Descartes, 75005 Paris, France}
\address[id=aff9]{Grupo de Estudios en Heliof{\'i}sica de Mendoza, Facultad de Ingenier{\'i}a, Universidad de Mendoza, 5500 Mendoza, Argentina}
\address[id=aff10]{Consejo Nacional de Investigaciones Cient{\'i}ficas y T{\'e}cnicas, C1425FQB Ciudad Aut{\'o}noma de Buenos Aires, Argentina}

\runningauthor{Joshi et al.}
\runningtitle{Multi-Ribbon Flare and Failed Filament Eruption}

\begin{abstract}
How filaments form and erupt are topics about which solar researchers have wondered since more than a century and that are still open to debate. We present observations of a filament formation, its failed eruption, and the associated flare (SOL2019-05-09T05:51) that occurred in active region (AR) 12740 using data from 
the Solar Dynamics Observatory (SDO), 
the Solar-Terrestrial Relations Observatory A (STEREO-A), the Interface Region Imaging Spectrograph (IRIS), and the Learmonth Solar Observatory (LSO) of the National Solar Observatory/Global Oscillation Network Group (NSO/GONG). AR 12740 was a decaying region formed by a very disperse following polarity and a strong leading spot, surrounded by a highly dynamic zone where moving magnetic features (MMFs) were seen constantly diverging from the spot. Our analysis indicates that the filament was formed by the convergence of 
fibrils at a location where magnetic flux cancellation was observed. Furthermore, we conclude that its destabilization was also related to flux cancellation associated to the constant shuffling of the MMFs. 
A two-ribbon flare occurred associated to the filament eruption; however, because the large-scale magnetic configuration of the AR was quadrupolar, two additional flare ribbons developed far from the two main ones. We model the magnetic configuration of the AR using a force-free field approach at the AR scale size. This local model is complemented by a global potential-field source-surface one.
Based on the local model, we propose a scenario in which the filament failed eruption and flare are 
due to two reconnection processes, one occurring below the erupting filament, leading to the two-ribbon flare, and another one above it between the filament flux-rope configuration and the large-scale closed loops. 
Our computation of the reconnected magnetic flux added to the erupting flux rope, compared to that of the large-scale field  overlying it, lets us conclude that the latter was large enough to prevent the filament eruption. A similar conjecture can be drawn from the computation of the magnetic tension derived from the global field model. 

\end{abstract}
\end{opening}
\section{Introduction}
\label{sec:intro}
Solar filaments are clouds of cool and dense plasma suspended against gravity by forces thought to be of magnetic origin. Filaments appear in H$\alpha$, 
Ca {\sc ii} images as dark features on the disk and as bright loops at the limb; this is  well explained by absorption and emission mechanisms. Prominences are bright also in transition region lines (He {\sc ii} 304 \AA) mapping the prominence-corona transition region, but dark in  some extreme ultraviolet (EUV) filtergrams due to continuum photoionization phenomena, {\it e.g.}  Fe {\sc xi} 171 \AA\ \citep{Labrosse2010}.
The main plasma characteristics of prominences are reviewed in \inlinecite{Labrosse2010}, while their magnetic properties are discussed in the articles by \inlinecite{Mackay2010} and \inlinecite{Gibson2018}. 

Prominences form along the magnetic polarity inversion line (PIL) in or between active regions. Early observations already suggested that their fine structure is apparently composed by many horizontal and thin dark threads \citep{Leroy1983, Bommier1986, Tandberg-Hanssen1995},  as it has been confirmed by observations using several telescopes, {\it i.e.}, the Télescope Héliographique pour l’Etude du Magn\'etisme et des Instabilités Solaires (THEMIS) \citep{Lopez2006, Schmieder2014, Levens2016}, the Solar Optical Telescope (SOT) on the Hinode satellite \citep{Berger2008}, and the New Vacuum Solar Telescope \citep[NVST,][]{Shen2015}. 
Some fine nearly horizontal plasma structures, lying in magnetic dips above parasitic polarities located in the filament channel, form the filament feet or barbs while the  endpoints  are anchored in the background magnetic field \citep{Lopez2006}. The  distance between these feet has a characteristic length comparable to the size of supergranules (30 Mm).
Even if prominences appear sometimes as hanging vertically over the limb their global structure is almost horizontal \citep{Martin1998,Chae2008}. Dynamics and projection effects could be responsible of such non-real appearance \citep{Schmieder2017}.  

Magnetic field extrapolations and magnetohydrodynamics (MHD) models have confirmed that the global structure of prominences consists of flux tubes or arcades of twisted magnetic field lines which have shallow dips in which cool plasma  is trapped \citep{Aulanier1998,vanBallegooijen2004}. In this aspect, prominences can be the cores of coronal-mass-ejection (CME) flux ropes  \citep{Fan2015} and their eruptions are the drivers of flares \citep{Devi2021}, in general, of the two-ribbon type \citep[see e.g. the standard model of flares discussed in][]{Aulanier2010,Schmieder2013}.

The review by \inlinecite[][and references therein]{Mackay2010} discusses the formation mechanisms of prominences. Different models are proposed based on levitation, evaporation and condensation processes. More recently, \inlinecite{Gibson2018} describes the formation of prominences and the structure of the magnetic skeleton that supports and surrounds the prominence, as well as how the plasma and magnetic field dynamically interact. 
Magnetic reconnection  between short filaments or chromospheric fibrils, sometimes accompanied by bidirectional jets \citep{Tian2017, Shen2017,Ruan2019,Shen2021}, may lead to the formation of long filaments \citep{Schmieder2004,Schmieder2006,Wang2007}; when this process happens close to parasitic polarities it may favour the formation of barbs. Such magnetic configurations correspond to the models proposed by \inlinecite{vanBallegooijen1989}.

High-resolution observations of coronal jets, mostly of the blow-out kind, have identified the presence and eruption of small-scale filaments, called minifilaments, as being part of the ejected material \citep{Hong2011,Shen2012,Sterling2015,Sterling2016,Panesar2017,Yang2018,Moore2018,Shen2019}. In another example, based on the analysis of lower resolution observations, the presence of a constantly reformed minifilament and its eruption was proposed as the origin of a series of blow-out jets and the chain of events following them \citep[flares and narrow CMEs,][] {Chandra2017jet}.  
The mechanism associated to the destabilization of the minifilament, as also happens with well-developed filaments, was the cancellation of magnetic flux along the polarity inversion line (PIL). Magnetic reconnection below the minifilament was responsible for an observed flare, while the same process above the minifilament favoured the injection of its material into open field lines to form the blow-out jet. The identification of minifilament eruptions as the main origin of the plasma ejected in these jets led \citet{Wyper2017} and \citet{Wyper2018} to propose that these ejections are produced by a break-out mechanism similar to the one proposed to explain larger events like CMEs \citep[see ][]{Karpen2012}. Several articles have reviewed different explanations (magnetic flux emergence and cancellation) for the origin of standard and blow-out jets using imaging and spectroscopic observations \citep[see e.g. ][and references therein]{Shen2021,Schmieder2022b}.

In general, not all eruptions end in a CME; there are partial and failed eruptions. A number of flux ropes and the embedded prominences suffer the later kind of ejections, which imply that at first they suddenly start to ascend, then decelerate, and stop raising at some larger height in the corona. Several cases of failed eruptions have been reported in the literature \citep{Shen2012,Chen2013,Joshi2013, Cheng2015,Thalmann2015,Xue2016,Chandra2017twostep,Nistico2017,Liu2018,Filippov2020,Filippov2021}.
\citet{Chen2013} and \citet{Xue2016} interpreted an unsuccessful eruption
because of  the presence of strong closed overlying EUV arcades. An asymmetry of the background magnetic field, considering only the relative location of the filament, has been suggested as the origin of failed eruptions \citep{Liu2009L}. \inlinecite{Joshi2013} studied the event of 17 June 2012; they discussed that the eruption of the flux rope and its filament could fail even after they reached up to the Large Angle and Spectrographic Coronagraph (LASCO) C2 field of view (FOV) and  were visible as a CME. These authors associated the failed CME to an asymmetric filament eruption. \citet{Thalmann2015} concluded that the strong overlying magnetic field over the active region (AR) 12192 in October 2014 prevented any CME to occur associated to  X-class flares . 
A comparative study of eruptive and non-eruptive events was performed by \inlinecite{Liu2018}. These authors explained non-eruptive events proposing two possibilities: firstly, the active region non-potentiality and a weak Lorentz force could be responsible for the small momentum of the ejecta and, secondly, the torus-stability region confined the eruption \citep[see ][for a discussion on the role of the torus instability]{Torok2005,Zuccarello2016}.  
Very recently \inlinecite{Filippov2021} estimated the mass of fifteen failed eruptive prominences using the model of a partial current-carrying torus loop anchored to the photosphere. Based on these calculations, they concluded that the gravity force could be the most suitable agent to stop the filament eruption. 
On the other hand, based on simulations, the articles by \inlinecite{Fan2003,Amari2018} propose  a simple  solution, i.e. a flux rope and embedded filament  do not erupt because of the overlying field that \citet{Amari2018} call a magnetic cage. 

\begin{figure*}[ht!]
\centering
\includegraphics[width=\textwidth]{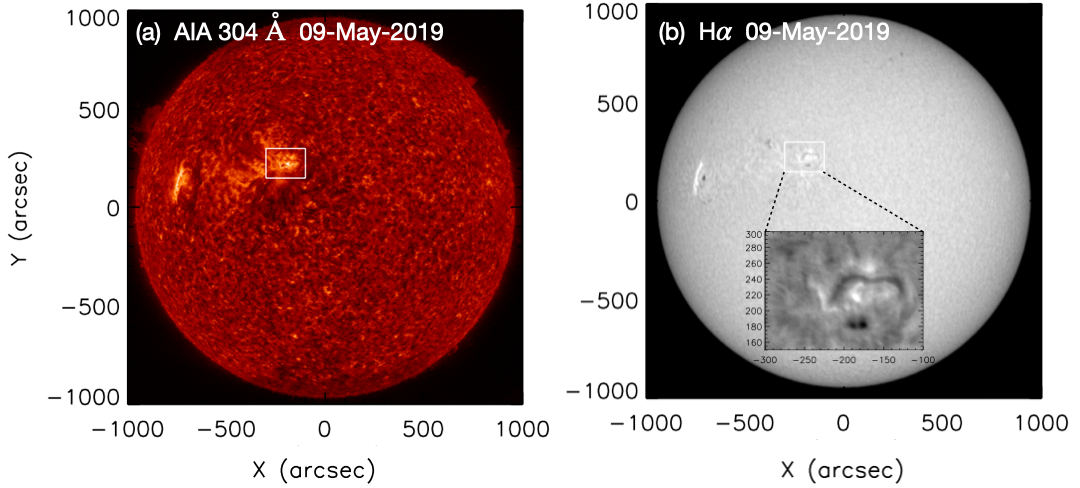}
\caption{Full-disk images showing AR 12740 ({\it white box}) in: (a) AIA 304 \AA\ band and (b) in GONG Learmonth H$\alpha$ image on 09 May 2019 at 05:39 UT including a zoom on the filament.  
The {\it white box} in panel b
covers the FOV of 
Figs. \ref{formation} and \ref{AIA}.}
\label{fulldisk}
\end{figure*}

In this article we present ground and space-based observations (\sect{data}) of a sequence of events (\sect{events}) that ended with the failed eruption of a filament. The chain of events (filament formation, failed eruption, and associated flare) occurred on 9 May 2019 in the decaying AR 12740, where the main sunspot was surrounded by a moat region, as well as several small bipole emergences. Consequently, we observe locations of 
emerging and cancelling flux 
leading first to the filament formation (\sect{filform}) and later to its eruption (\sect{filerup}). The eruption, which failed, was accompanied by a flare (\sect{flare}) of C6.7 X-ray class recorded by the Geostationary Operational Environmental Satellite (GOES) starting at 05:40 UT, a maximum at 05:51 UT, and an extension of around two hours. Figure~\ref{fulldisk} shows AR 12740 in full disk images at the time of the flare. We present local and global magnetic field models in \sect{model} and, based on our modeling and observations, we propose a scenario to explain the observed events (\sect{scenario}). 
Finally, we summarize and conclude in \sect{discussion}.

\section{The Data Used}
\label{sec:data}

To analyse the series of events that occurred in AR 12740 on 9 May 2019, we use extreme ultraviolet (EUV) and ultraviolet (UV) data from the Atmospheric Imaging Assembly \citep[AIA:][]{Lemen2012}, on board the Solar Dynamics Observatory (SDO), EUV observations from the Extreme Ultraviolet Imager \citep[EUVI:][]{Wulser04} of the Sun-Earth Connection Coronal and Heliospheric Investigation suite \citep[SECCHI:][]{Howard08}, on board the Solar-Terrestrial Relations Observatory (STEREO) spacecraft A,  and from the Interface Region Imaging Spectrograph \citep[IRIS:][]{DePontieu2014}. H$\alpha$ data come from the Learmonth Solar Observatory (LSO) of the National Solar Observatory/Global Oscillation Network Group (NSO/GONG) and magnetograms from the Helioseismic and Magnetic Imager \citep[HMI:][]{Scherrer2012}, on board SDO.  

AIA provides full-disk images at seven EUV and two UV wavebands, with a pixel size of 0.6$\arcsec$~and a cadence of 12 s and 24 s for EUV and UV, respectively.
The higher temperature wavebands, including 94 \AA\ (6.3 MK), 131 \AA\  (0.40 MK, 10 MK, 16 MK), 171 \AA\ (0.63 MK), 193 \AA\ (1.3 MK, 20 MK), 211 \AA\  (2.0 MK), and 335 \AA\ (2.5 MK), typically show features in the corona such as loops. The lower temperature wavebands, 304 \AA\ (0.050
MK), 1600 \AA\ (0.10 MK), and 1700 \AA\ (continuum) are sensitive to heating in the chromosphere. 
In our analysis we use 304 \AA, 171 \AA, and 1600 \AA~bands (henceforth, AIA 304, AIA 171, and AIA 1600). We select, from the full-disk images, subimages containing the region of interest. The images are coaligned to compensate for solar rotation and the movies that accompany this article are constructed (\sect{flare}). The images are either displayed in  logarithmic intensity scale for better contrast or using the multi-scale Gaussian normalization \citep[MGN:][]{Morgan2014} processing technique.

We complement the SDO/AIA data with full-disk observations in the 304 and 195 \AA~channels of the STEREO-A/EUVI instrument (henceforth, EUVI-A 304 and EUVI-A 195). EUVI provides images with a pixel size of 1.6$\arcsec$ and a temporal cadence of 10 minutes for EUVI-A 304 and 5 minutes for EUVI-A 195 during the analysed events. On 09 May 2019, the STEREO-A spacecraft was located at an Earth ecliptic (HEE) longitude of -95$\degree$; from this location AR 12740 was seen at the west solar limb. 

IRIS observed AR 12740 between 04:54 UT and 06:21 UT in the mode of very dense rasters and, simultaneously, obtained slit jaw images (SJIs)  centered on the AR with a FOV of 167  $\times$ 175  in four channels around  1330 \AA,  1400 \AA, 2796 \AA, and 2832 \AA, including C \textsc{ii}, Si \textsc{iv}, Mg \textsc {ii} lines, and the UV continuum, respectively. 
C \textsc{ii} is formed around T = 30000 K and Si \textsc{iv} around 80000 K, while Mg \textsc{ii} is formed at chromospheric temperatures between 8000 K and 20000 K. The cadence of the SJIs is 65 sec and the pixel size is 0.35$\arcsec$. 

The H$\alpha$ data come from LSO and have a spatial resolution of approximately 2$\arcsec$; they are obtained with a cadence of 1 min. The analysed SDO/HMI data consist of line-of-sight (LOS) full-disk magnetograms (0.5$\arcsec$ pixel size) and synoptic maps. As done for AIA, we select from the full-disk magnetograms subimages centered in the AR and, after coalignment, we construct the movies that are attached to this article (\sect{obs_B}). The magnetograms are used to study the evolution of the AR magnetic field, as described in \sect{obs_B} (with 45 s cadence), and as boundary condition for the local model described in \sect{local_model} (720 s cadence). HMI  synoptic  maps  are  computed  from  LOS  magnetograms  by  combining central meridian data from 20 magnetograms collected along a 4-hour interval each day. A  synoptic  map  is  made with  the  magnetograms  observed  over  a  full  solar  rotation with  3600 $\times$ 1440  steps in  longitude  and  sine  latitude. Details  concerning  the  construction  of  synoptic  maps  can  be  found  in  the HMI web-site \url{jsoc.stanford.edu/jsocwiki/SynopticMaps}; the map for Carrington rotation (CR) 2217 is used as boundary condition for the model in \sect{global_model}. 

\begin{figure*}[ht!]
\centering
\includegraphics[width=\textwidth]{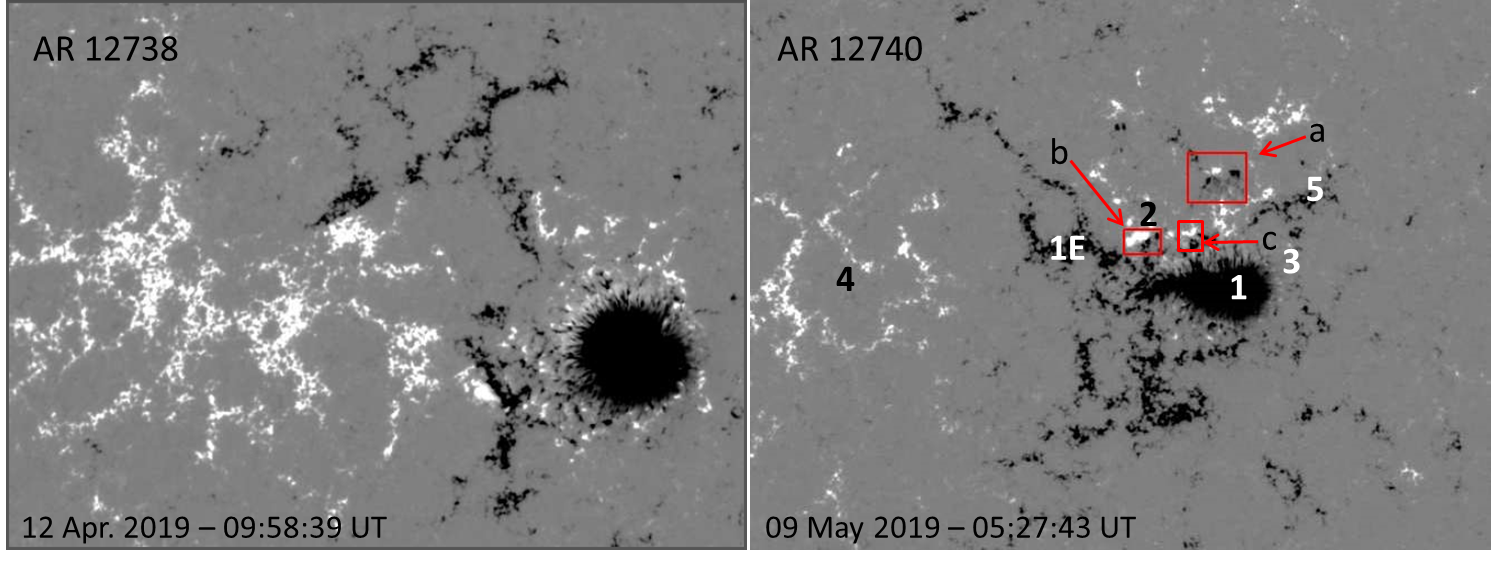}
\caption{{\it Left panel}: Large scale magnetic configuration of AR 12738 on April 2019 on CR 2216. A compact preceding negative spot, surrounded by a moat region, is followed by a disperse following positive polarity. {\it Right panel}: AR 12740, the return of AR 12738 on the next CR, 
showing a similar configuration. The {\it red rectangles} surround regions where we observe magnetic flux cancellation probably related to the events that occurred on 9 May 2019; they are pointed with {\it arrows} and labelled as \texttt{a}, \texttt{b}, and \texttt{c}. Different magnetic polarities (or their extensions) that are relevant to our study are indicated with {\it numbers} (or a {\it number} and a {\it letter}). 
In both panels, {\it white (black)} regions correspond to positive (negative) LOS magnetic field measurements. The magnetic field values have been saturated above (below) 300 G (-300 G).
The size of each panel is 330$\arcsec$ in the E–W (east–west) and 244$\arcsec$ in the N-S (north-south) direction. The center of each panel in heliographic coordinates is N06 E07 for the {\it left panel} and N08 E01 for the {\it right panel}. A movie covering the evolution of AR 12740 from 7 to 9 May 2019 accompanies this figure \url{(HMI_7-9May2019_Fig2.mp4}); the magnetic field values have been saturated above (below) 500 G (-500 G) for a better visualization of bipole emergences and changes along these days.}
\label{twoCRs}
\end{figure*}

\section{The Events on 9 May 2019 in AR 12740}
\label{sec:events}

\subsection{The Magnetic Field Evolution}
\label{sec:obs_B} 
AR 12740 appeared on the eastern solar limb on 4 May 2019. By the time of the events described in this article it was located at N10 E07. This AR is the return of AR 12738 on the previous CR.  Figure~\ref{twoCRs} shows the magnetic field distribution on 12 April and 9 May 2019. On 12 April, AR 12738 consisted of a leading concentrated negative polarity spot already in its decaying phase followed by a disperse positive polarity to the east. An extended moat region was present around the main negative polarity. Moat regions \citep[see][and references therein]{vanDriel2015}, which appear mostly around evolved and decaying spots play a key role in transporting flux away from spots and, therefore, contributing to their decay. Furthermore, moat regions are the sites of active phenomena, e.g. eruptions and recurrent jets \citep{Chen2015,Chandra2015,Shen2018}. The moat region around the strong negative spot in AR 12738 was also present one rotation later (compare both panels in Fig.~\ref{twoCRs}).

\begin{figure*}[ht!]
\centering
\includegraphics[width=\textwidth]{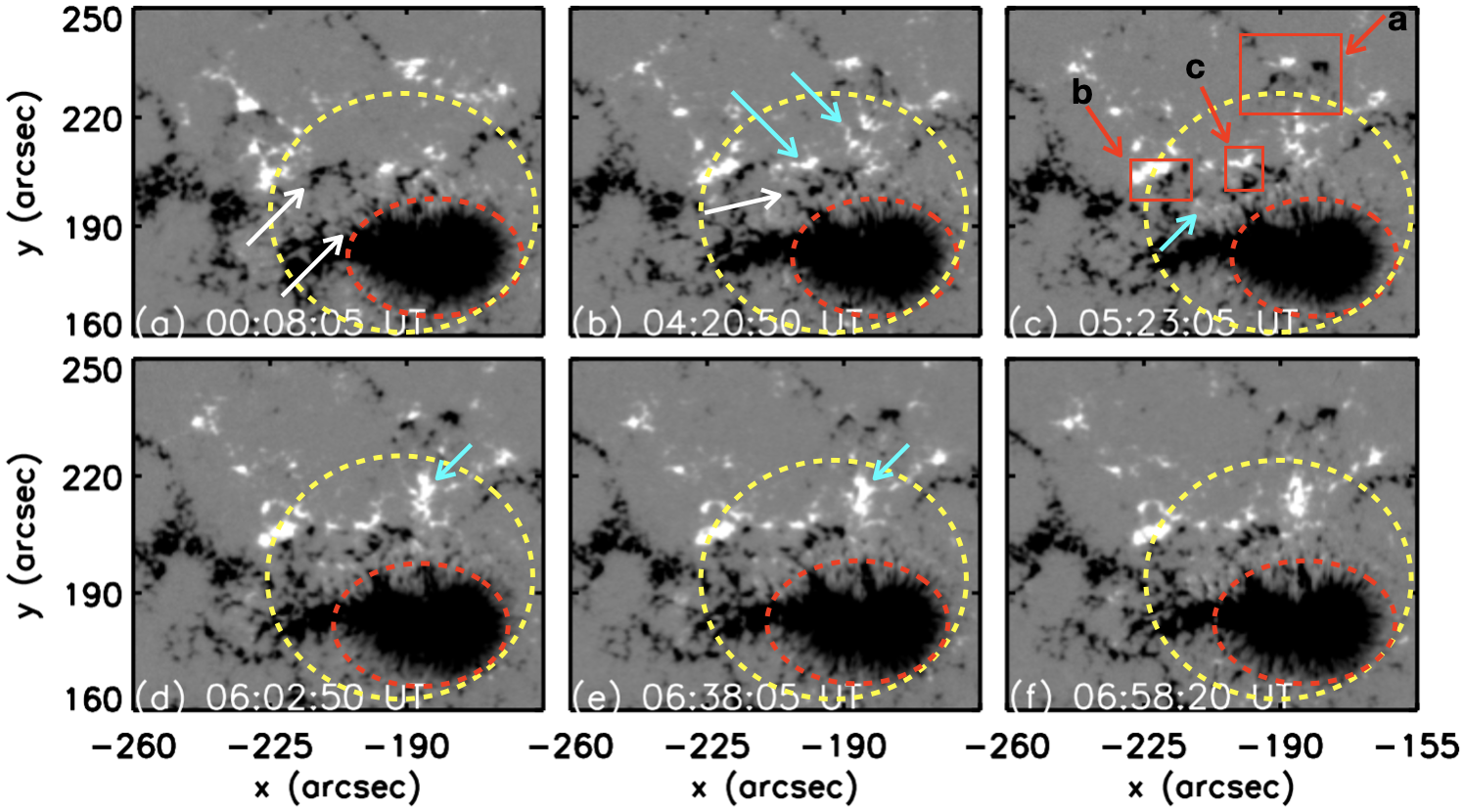}
\caption{Evolution of part of the moat region surrounding the leading negative spot between 00:08:05 UT and 06:58:20 UT on 9 May 2019. {\it White/cyan arrows} indicate negative/positive MMFs rapidly changing. The {\it red oval} and the {\it yellow circle} have the same size in all the panels; this facilitates the visualization of the contraction of the main negative polarity and the expansion of the region where MMFs are visible.  In panel (c) red boxes \texttt{a}, \texttt{b}, \texttt{c}, similar to those in Fig.~\ref{twoCRs}, are drawn and indicated by {\it red arrows}. The magnetic field values have been saturated above (below) 400 G (−400 G). A movie with a similar FOV and of similar saturation accompanies this figure \url{(HMI_09May2019_Fig3.mp4)}.}
\label{MMF}
\end{figure*}

The evolution of the moat region is well visible in the panels presented in Fig.~\ref{MMF} and the accompanying movie. This figure shows the leading negative spot (red oval) surrounded by a part of the moat region (yellow circle). 
The main spot decreases in size while small magnetic features, called moving magnetic features \citep[MMFs:][]{Harvey1973} move away from the spot (see arrows).

Besides this constant radial motion of the MMFs,  we observe the emergence of several small bipoles toward the north of the main spot that made the full configuration highly dynamic (see the movie \url{HMI_09May2019_Fig3.mp4}). These series of emergences and their consequent evolution created a PIL nearly E-W oriented where a filament formed as discussed in \sect{filform}. 

\begin{figure*}[t!]
\centering
\includegraphics[width=\textwidth]{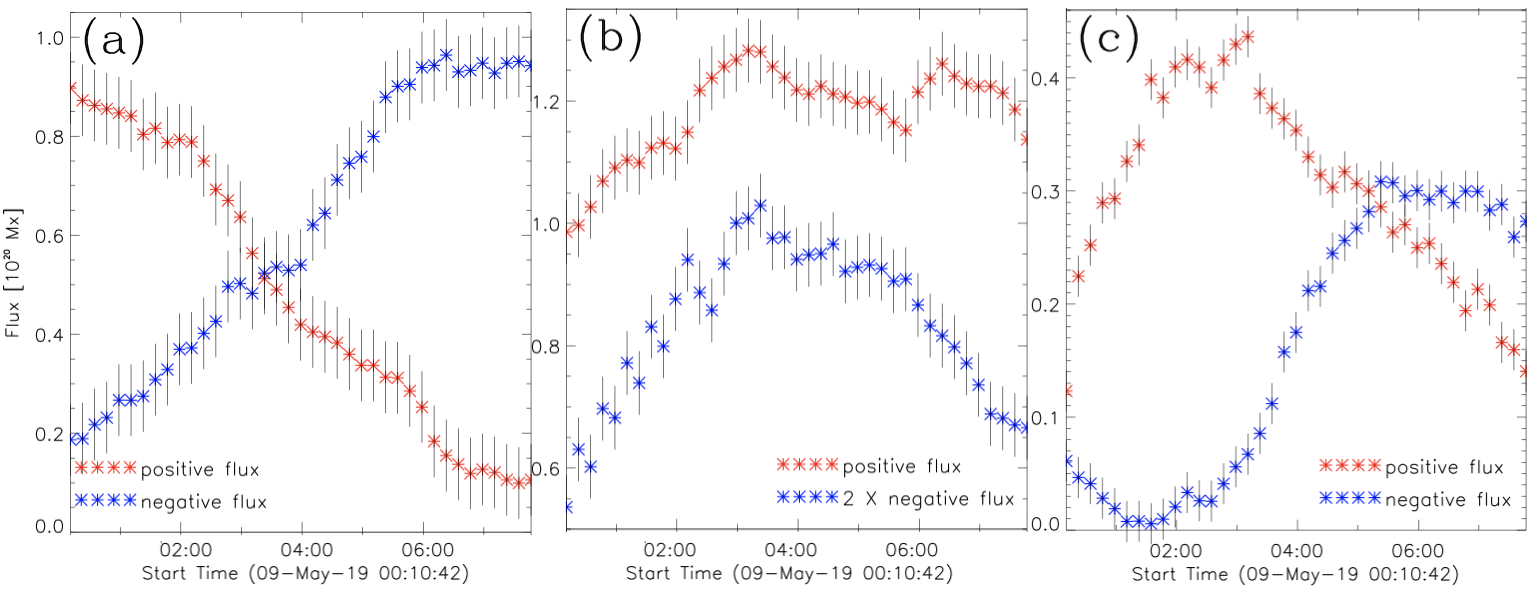}
\caption{Evolution of the positive and absolute value of the negative magnetic fluxes in regions related to the filament formation and eruption. (a) Corresponds to the region within the {\it rectangle} labelled as \texttt{a} in Fig.~\ref{twoCRs} where two long and wide fibrils merge to form the curved AR filament. (b) and (c) Show the flux evolution in regions labelled as \texttt{b} and \texttt{c} in Fig.~\ref{twoCRs} that can be associated to the filament eruption. Notice that in panel b the {\it blue asterisks} are multiplied by 2. Computations are done for values of the field above $10$~G and the error bars are calculated considering a magnetic field error of 5~G.}
\label{cancelflux}
\end{figure*}

We also identify several locations where flux cancellation occurred. Some of them are relevant to either the filament formation or its destabilization as discussed in \sects{filform}{filerup}, {\it i.e.} see the rectangular boxes in Fig.~\ref{twoCRs} (right panel) pointed with arrows and labelled as \texttt{a}, \texttt{b}, and \texttt{c}. Figure~\ref{cancelflux} shows the evolution of the positive and negative magnetic fluxes within these boxes. Panel a corresponds to the region (labelled as \texttt{a} in Fig.~\ref{twoCRs}) where we identify the merging of two elongated and wide fibrils that finally formed a curved filament (see Fig.~\ref{formation} and \sect{filform}); notice that only the positive flux is seen decreasing while the negative flux increases as it enters the southern boundary of this northern box. Panel b corresponds to the region (labelled as \texttt{b} in Fig.~\ref{twoCRs}) where we start observing the development of the two main ribbons of the C6.7 flare (see \sect{flare}); notice that in this case both negative and positive fluxes steadily decrease from around 03:10 UT until around 05:50 UT. Panel c shows the flux evolution in the rectangle (labelled as \texttt{c} in Fig.~\ref{twoCRs}). As in the case of region \texttt{a}, only the positive flux is seen decreasing after around 03:10 UT because negative flux, advected by the moat flow, enters the southern border of this rectangle. This region (\texttt{c}) could be related to the filament destabilization (see the discussion in \sect{flare}).

By the beginning of the flare and filament eruption (see \sects{flare}{filerup}), the magnetic field distribution is the one depicted in the right panel of Fig.~\ref{twoCRs}.
Since the magnetic configuration and its evolution is complex, we first limit its description to the quadrupolar configuration relevant for the studied flare and filament eruption. This involves  polarities 1, 2, 3, and 4. 
Polarities 1 and 4 are the main ones of the AR. Polarity 2, which dramatically evolves in the hours previous to the flare, adds up to the quadrupolar layout. The fourth polarity that we call 3 is located to the west of polarity 1; following the evolution of the moat region around the main AR negative spot, this chain of polarities is formed by the MMFs moving away from the big spot. 

In Fig.~\ref{twoCRs}, we have also labeled the extension of polarity 1, which ends at the border of a supergranular cell to the east, as 1E, as well as a north-western negative polarity that we call 5. This polarity is part of a bipole that emerged as early as 7 May 2019 at around 20:50 UT and evolved to the position shown in Fig.~\ref{twoCRs} on 9 May; the positive bipole polarity is located to its north. Both 3 and 5 serve as a reference for our discussion in \sect{model}. A movie displays the complex evolution  of AR 12740 from early 7 May to 9 May after the flare decay \url{(HMI_7-9May2019_Fig2.mp4}).

\subsection{The Filament Formation} 
\label{sec:filform} 

\begin{figure*}[ht!]
\centering
\includegraphics[width=\textwidth]{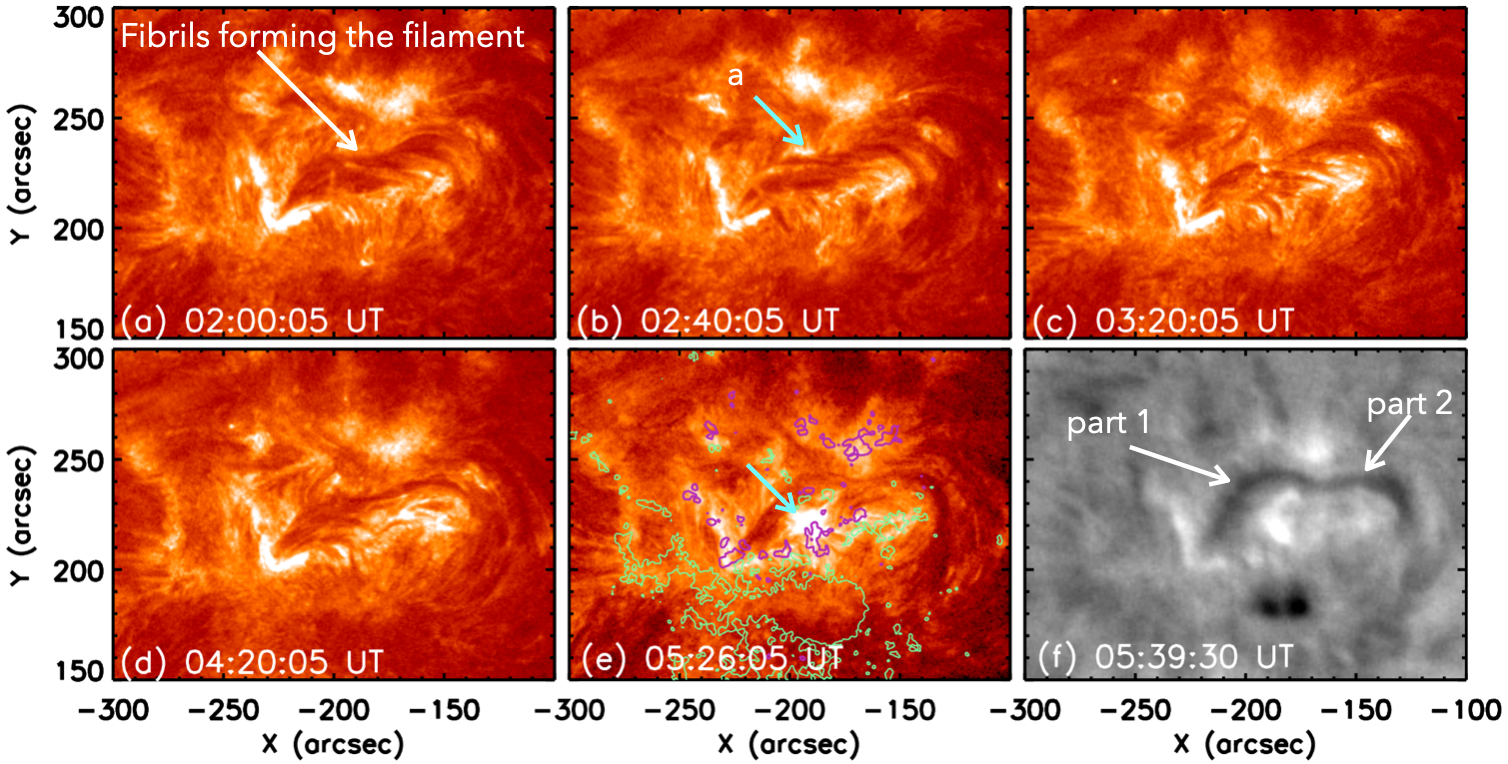}
\caption{Formation of a long filament by the merging of fibril arcades to the N and W of the main sunspot in AR 12740 observed in AIA 304 (panels a-e) on 9 May 2019. Flux cancellation occurred in site \texttt{a} around 02:40 UT and continue (see {\it arrows} in panels a and b and the discussion in the text); this favored the formation of a long curved structure. See text for the description of the evolution of this structure and the appearance of minor brightenings. A movie extending from 01:00 UT to 07:00 UT on 9 May accompanies this figure and Fig.~\ref{AIA} (\url{AIA304_09May2019_Fig3_Fig6.mp4}). HMI contours of $\pm$100 Gauss ({\it magenta/green} for positive/negative polarities) are overlaid  in panel e. 
Panel f presents H$\alpha$ observations of the filament one minute before flare onset; its two parts are labelled as part 1 and part 2 and are pointed with  {\it white arrows} (see text).} 
\label{formation}
\end{figure*}

A long and curved filament started forming a few hours before the flare initiation time at around 05:40 UT. Figure~\ref{formation}a shows sets of very long, wide, and winding fibrils at 02:00 UT in AIA 304. 
Part of these fibrils were involved in the merging process to form the long filament (Fig.~\ref{formation}d).
These fibril arcades  evolved as time went on and seemed to merge at the location of magnetic flux cancellation; the white arrow in panel a points approximately to the magnetic flux cancellation site called \texttt{a} in Fig.~\ref{twoCRs} (right panel), whose evolution is shown in Fig.~\ref{cancelflux}a. Panels b and c of Fig.~\ref{formation} depict this evolution. However, because the flux cancellation process was accompanied by minor brightenings (see the light blue arrow in panel b), the elongated and curved filament structure appears interrupted by them as can be better seen in panel c. 
We have called part 1 (labelled in Fig.~\ref{formation}f) to the eastern fibril arcade. Its negative polarity footpoints lie on the negative polarity at the flux cancellation site \texttt{a}. Its other footpoints are anchored in the E-W branch of the positive polarity called 2 in Fig.~\ref{twoCRs} (right panel). Notice that polarity 2 has a global L-shape, with the longest part of the L in the N-S direction and the shortest in the E-W direction. Furthermore, by the time of all panels in Fig.~\ref{formation} an L-shape plage brightening is seen tracing the polarity global shape.  On the other hand, we have called part 2 (labelled in Fig.~\ref{formation}f) to the western fibril arcade with positive polarity footpoints 
at site \texttt{a} and negative footpoints most probably anchored in the disperse negative polarity 3 to the west of the leading spot (see Fig.~\ref{twoCRs}, right panel). 
By around 04:20 UT (see Fig.~\ref{formation}d), the filament appeared as a single elongated and curved structure following the complex PIL created by the dynamics of the constant shuffling of the MMFs. However, by around 05:26 UT (see Fig.~\ref{formation}e), the filament appears again as separated in two parts because of a brightening associated to the flux cancellation site called \texttt{c} in Fig.~\ref{twoCRs} (right panel). When seen in a high-time resolution movie in AIA 304, this bright kernel marked by a light blue arrow in Fig.~\ref{formation}e, corresponds to a small and localized jet and is not associated to the main C6.7 flare. Later, by 05:39 UT, one minute before flare onset in GOES, the filament is seen as a long curved structure in H$\alpha$ (see panel f). 

In summary, the filament is associated to opposite polarities converging and cancelling. This builds up progressively a coherent structure. Our observations agree and add up to previous studies. Indeed, the evolution of fibrils merging and forming a filament has been already observed in cases where a filament formed from a loop arcade \citep[see e.g.][]{Guo2010}. 
Furthermore, the basic process of flux cancellation at fibril footpoints creating long  magnetic field lines is also well described by \citet{vanBallegooijen1989} and \citet{Schmieder2004}.
Filament formation from magnetic reconnection between adjacent short filament threads was observed and analysed in EUV and H$\alpha$ observations \citep{YangBo2016,XueZhike2017,ChenHechao2018}.
 This kind of merging of short threads or fibrils through magnetic reconnection can originate bi-directional jets along the newly formed structure \citep{Tian2017, Shen2017}. We also observe these jets in an IRIS spectrum movie of our case study; however, it is out of the scope of this article to analyse IRIS spectra, we just add that bi-directional jets found in IRIS SJIs and spectra are well discussed in previous articles  \citep{Ruan2019,Joshi2021}.

\begin{figure*}[ht!]
\centering
\includegraphics[width=\textwidth]{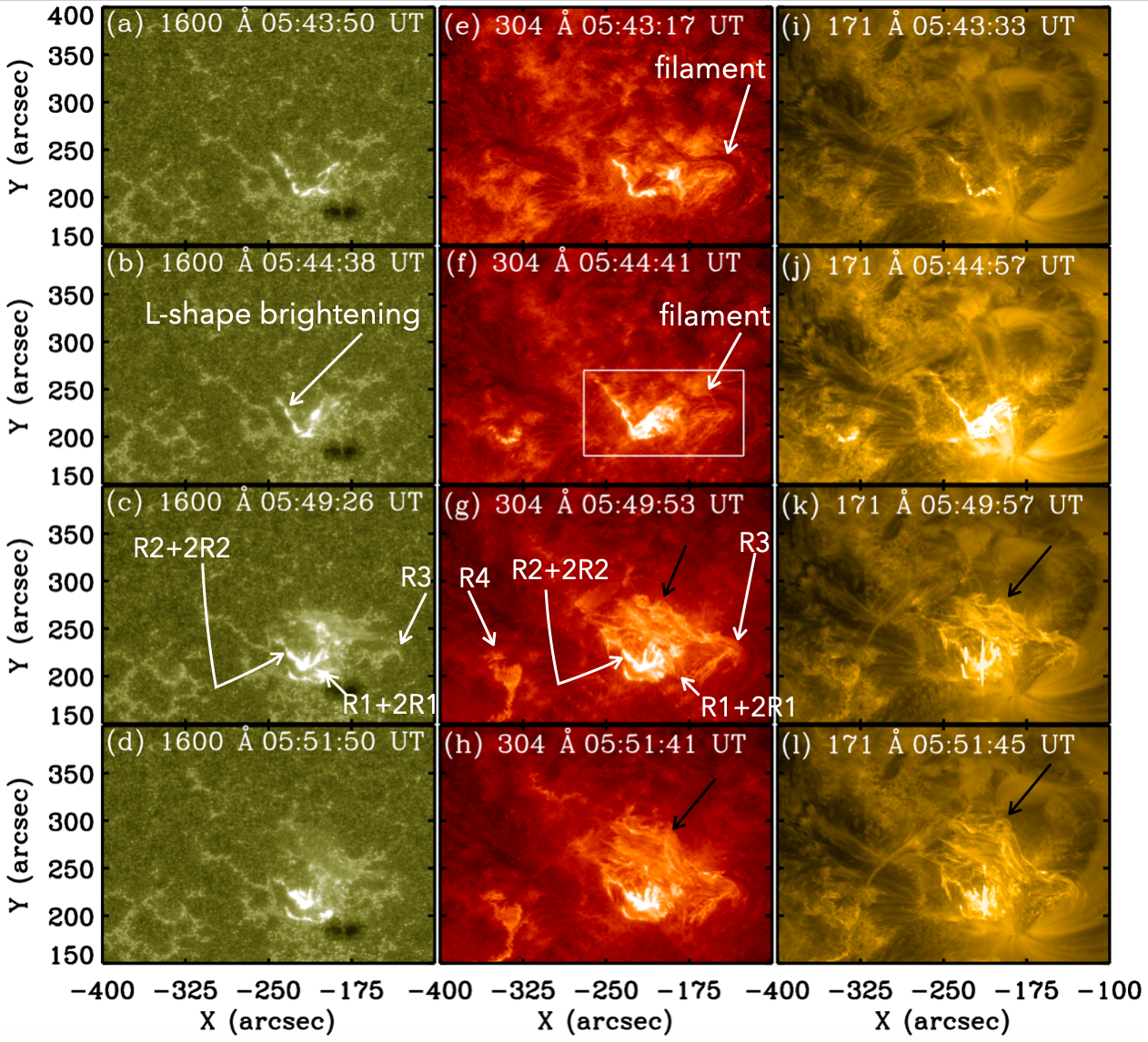}
\caption{Evolution of the flare and the filament eruption in AR 12740 observed on 9 May 2019. The {\it left column} corresponds to AIA 1600, the {\it middle column} to AIA 304, and the {\it right column} to AIA 171. The  {\it box} in panel f indicates the FOV of IRIS. The main flare ribbons visible in AIA 1600 are pointed with {\it arrows} in panel c and in AIA 304 in panel g. The western portion of the filament, which erupted few minutes after the eastern portion, is pointed with an {\it arrow} in panels e and f of AIA 304 images. {\it Black arrows} in panels g, h, k, and l indicate the northern edge of the heated filament plasma as it erupts. See text for the description of this figure and the movie \url{AIA304_09May2019_Fig3_Fig6.mp4}. }
\label{AIA}
\end{figure*}

\begin{figure*}[t!]
\centering
\includegraphics[width=\textwidth]{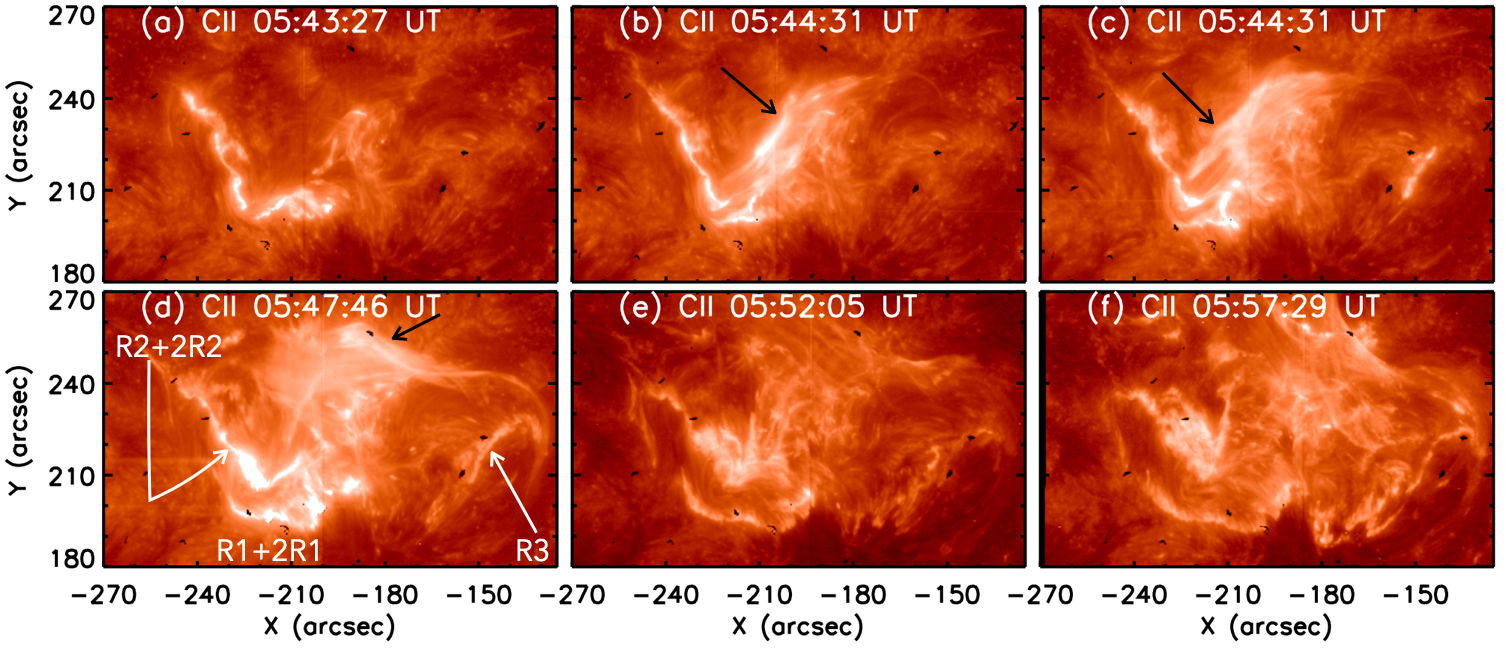}
\caption{Evolution of the flare and filament eruption observed with IRIS 1330 \AA\ channel between 05:43:27 UT and 05:57:29 UT. The main flare ribbons are pointed with {\it white arrows} and labeled in panel d. 
The evolution of the rising filament is indicated by {\it black arrows} in panels b, c, and d. The up-going heated plasma of the filament is pointed with an {\it arrow} in panel d, as well as the western flare ribbon R3 that is also visible at this time. This FOV is indicated in Figure \ref{AIA}f. See text for the description of this figure and the accompanying movie \url{IRIS_CII_09May2019_Fig7.mp4}.} 
\label{IRIS}
\end{figure*}

\subsection{The Flare and Its Multiple Ribbons}
\label{sec:flare}

The evolution of the C6.7 flare is shown in three AIA wavelength ranges  (AIA 1600 \AA, AIA 304 \AA, and AIA 171 \AA) in Fig.~\ref{AIA} and in IRIS 1330 \AA\ channel SJIs (Fig.~\ref{IRIS}); notice that AIA images depict a larger FOV than that of IRIS. 
Two movies with different temporal and spatial extensions accompany the figures in this section, \url{AIA304_09May2019_Fig3_Fig6.mp4} and \url{IRIS_CII_09May2019_Fig7.mp4}. In both, despite the saturation in several images, the evolution of the flare and filament eruption at all of their stages can be followed. 

Before describing the flare temporal evolution, we define the labeling of the distinctly observed flare ribbons in Figs.~\ref{AIA} and~\ref{IRIS} to facilitate our following description. This C6.7 flare consists mainly of a two-ribbon flare that occurred within the large quadrupolar magnetic configuration of the AR (see Fig.~\ref{twoCRs} right panel). The double ribbons of the two-ribbon flare are called 2R (for two-ribbon flare)  followed by a number that corresponds to the polarity number where the ribbon is located as shown in  Fig.~\ref{twoCRs}. We have also identified two additional ribbons, R3 located on the western and disperse negative polarity R3 and R4 located on polarity 4. These two ribbons are visible in  Figs.~\ref{AIA} and \ref{IRIS} and in the larger H$\alpha$ FOV in Fig.~\ref{Halphaevolution}. 

By around 05:43 UT the L-shape brightening, described in Sect.~\ref{sec:filform}, is the most evident feature in the three AIA bands (see Fig.~\ref{AIA}), from chromospheric to low coronal temperatures. The evolution of the two-ribbon flare starts along the E-W extension of this brightening. The separation of its main bands is clear and increasing as in a typical two-ribbon flare in panels c, d, g, and h of Fig.~\ref{AIA} of both AIA 1600 and AIA 304. The relative shift of these two ribbons along the PIL
indicates the presence of high magnetic shear at that location. Concerning IRIS, we focus on the C {\sc ii} band pass SJIs, in this band the two main ribbons are observed already at 05:43 UT in Fig.~\ref{IRIS}a because IRIS SJIs have higher spatial and spectral resolution than AIA images though a smaller FOV. Their evolution and separation is clearer than in Fig.~\ref{AIA}.

By around 05:49 UT, a ribbon that we label as R3 in panels c and g of Fig.~\ref{AIA} is present to the west of the FOV on polarity 3. Simultaneously, another very elongated brightening is clearly seen to the east in Fig.~\ref{AIA}g and h, we have labeled it as R4. In the higher temperature AIA band, AIA 171, the northern portion of R4 appears in Fig.~\ref{AIA}j and its shape can be guessed in panels k and l. In a similar way as with the main ribbons 2R1 and 2R2, R3 is better seen in Fig.~\ref{IRIS}c and d; however, R4 is not visible because of the reduced IRIS FOV. 

Based on the appearance of the distant ribbons, R3 and R4, and our magnetic field model in Sect.~\ref{sec:local_model}, we conclude that the counterparts of R3 and R4 should be located on polarities 1 and 2 but we are not able to separate them clearly from 2R1 and 2R2. That is why, we have labeled the extended ribbons along polarities 1 and 2 as R1+2R1 and R2+2R2 (see Figs.~\ref{AIA} and \ref{IRIS}) to indicate that they are possibly a combination of the main ribbons of the two-ribbon flare and the counterparts of R3 and R4 within the quadrupolar AR configuration. 

Another feature, better seen at $\approx$ 05:47 UT in Fig.~\ref{IRIS}d, is a curved brightening to the north of R3. This brightening is located on polarity 5 and the field line connectivity derived in Sect.~\ref{sec:local_model} lets us conclude that it is not related to the C6.7 flare. 

\begin{figure*}[t!]
\centering
\includegraphics[width=\textwidth]{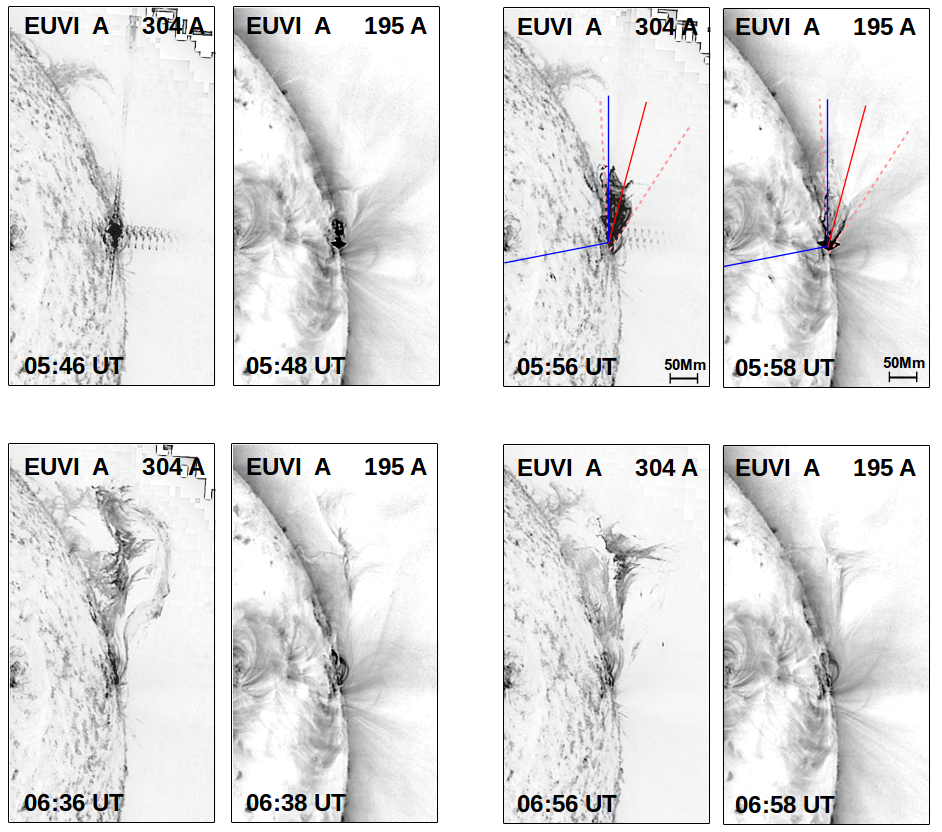}
\caption{STEREO-A/EUVI images of AR 12740 in the 304 and 195 \AA~channels shown side by side at  close-in-time hours. The spacecraft was approximately located on the ecliptic at an Earth ecliptic (HEE) 
longitude of $-95 \degree$ and the AR is observed on the solar limb. The saturated pixels ({\it in black}) correspond to the flare. The average direction of the mean prominence/plasma motion ({\it red}),  the radial direction ({\it blue}),
and the N-S direction ({\it blue}) are marked with {\it solid lines} in the top-right pair of panels. In the same panels, the {\it dashed-red} lines indicate the plasma ejection width projected in the plane-of-the-sky. A {\it segment} has been added to the bottom right to indicate the figure scale-size. The observation times are provided at the bottom left of each panel.}
\label{fig:stereo}
\end{figure*}

\subsection{Failed Eruption of the Filament}
\label{sec:filerup}

In this section we describe the different observed stages of the filament eruption, from its lift off to the return of its plasma after its eruption has failed. We first discuss the observations as seen at the solar limb by EUVI in STEREO-A because from them, we can derive the ejection direction to help understanding the eruption as seen from Earth's perspective.

\subsubsection{The Failed Eruption from STEREO-A Point of View}
\label{sec:failed_STEREO}

At the time of the event, STEREO-A was at a privileged location to observe the coronal activity related to AR 12740. From the STEREO-A point of view, AR 12740 appeared on its western  solar limb, as shown in \fig{stereo}. The panels of this figure correspond to EUVI-A 304 and 195 at different times from a few minutes after the beginning of the flare, when it is clearly seen on the limb of STEREO-A,  and cover the filament eruption and consequent observation of plasma downflows. In this figure pairs of panels at similar hours are shown side by side for both channels. The images in these panels are shown in a gray-reversed scale and have been processed using a wavelet transform \citep[see ][]{Stenborg2008}.  We point the reader to the movies that can be generated at \url{cdaw.gsfc.nasa.gov/stereo/daily_movies/} not only showing the EUVI-A low corona but also the white-light corona as imaged by COR1-A and COR2-A.

The vantage point of view of STEREO-A provides information on the failed eruption, which is inaccessible from Earth's line of sight. We have used EUVI-A 304 and 195 images to compute the angles that the N-S and radial  directions made with the average direction of the upflowing plasma. These angles are estimated from the pair of panels at the top right in \fig{stereo}, where the average direction of the plasma upflow is shown with a red solid line and the radial and the N-S directions with blue ones. The dashed-red lines indicate the plasma ejection width as projected on the plane-of-the-sky. The  measured angles are almost the same for both channels: $\approx 15\degree$ and $\approx 60\degree$ with respect to the N-S and radial directions, respectively. These values can be used to correct those of variables computed using data obtained from Earth's point of view (see \sect{failed_AIA} and \sect{global_model}).

We observed a CME whose leading edge appeared in STEREO-A COR1 FOV at 05:55 UT. However, we could not identify the CME  source region in AIA images. Furthermore, the probable CME launch time, considering its projected speed in COR1-A images, would be before the start of the C6.7 flare. Therefore, we conclude that this CME is not related to the filament eruption we study (see Appendix). 

\begin{figure*}[t!]
\centering
\includegraphics[width=\textwidth]{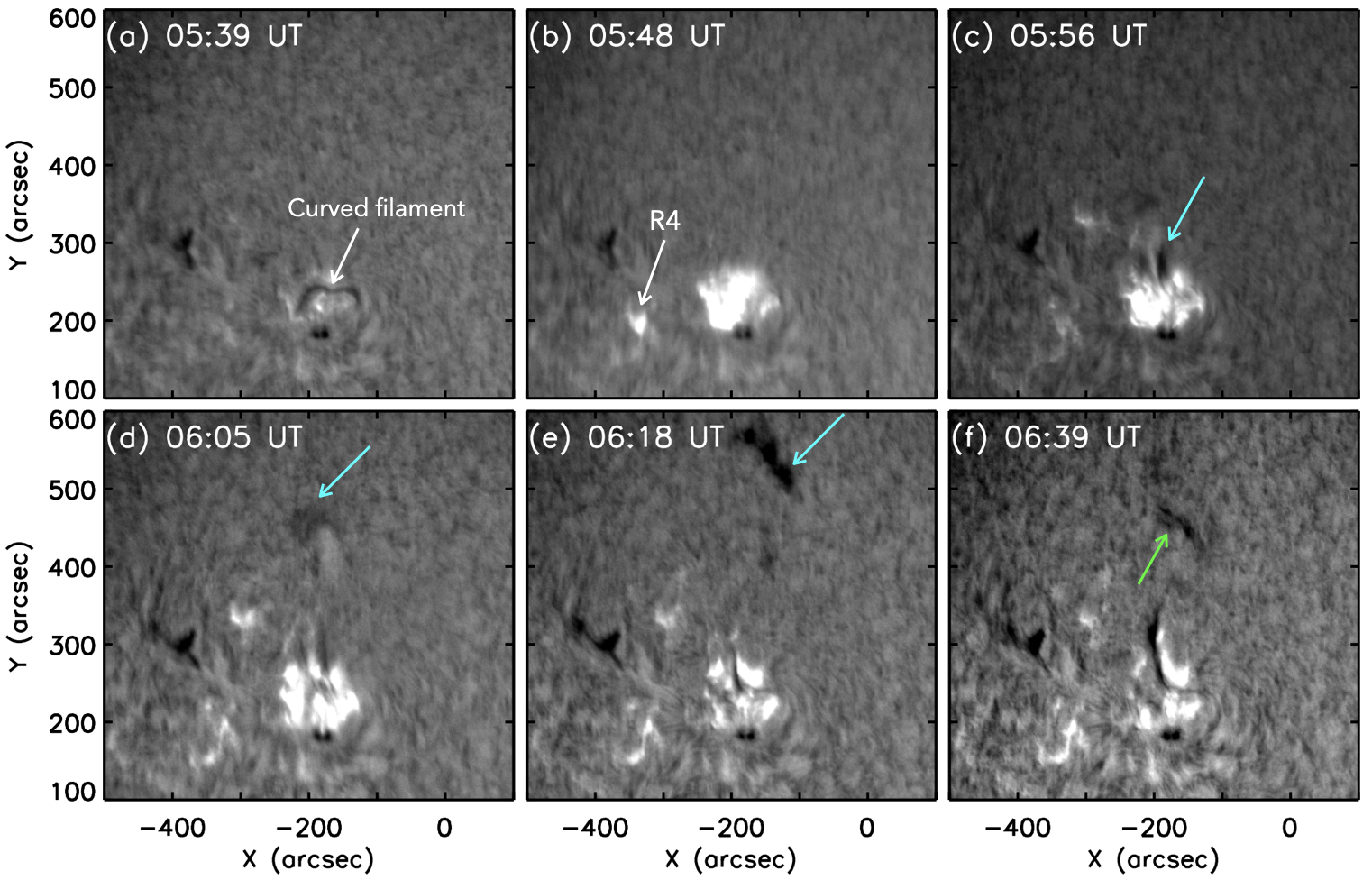}
\caption{Evolution of the eruption in H$\alpha$ observations stressing mainly its later stages. The curved filament before the eruption is well visible in panel a. In panel b we have labeled as R4 the elongated ribbon with a top rounded shape identified in AIA images (Fig.~\ref{AIA}). The cool material going upward is pointed with {\it light blue arrows} in panels c-e and with a {\it green arrow} when it is falling back in panel f. See the accompanying movie \url{(Halpha_09May2019_Fig8.mp4)}.}
\label{Halphaevolution}
\end{figure*}

\begin{figure*}[ht!]
\centering
    \includegraphics[width=\textwidth]{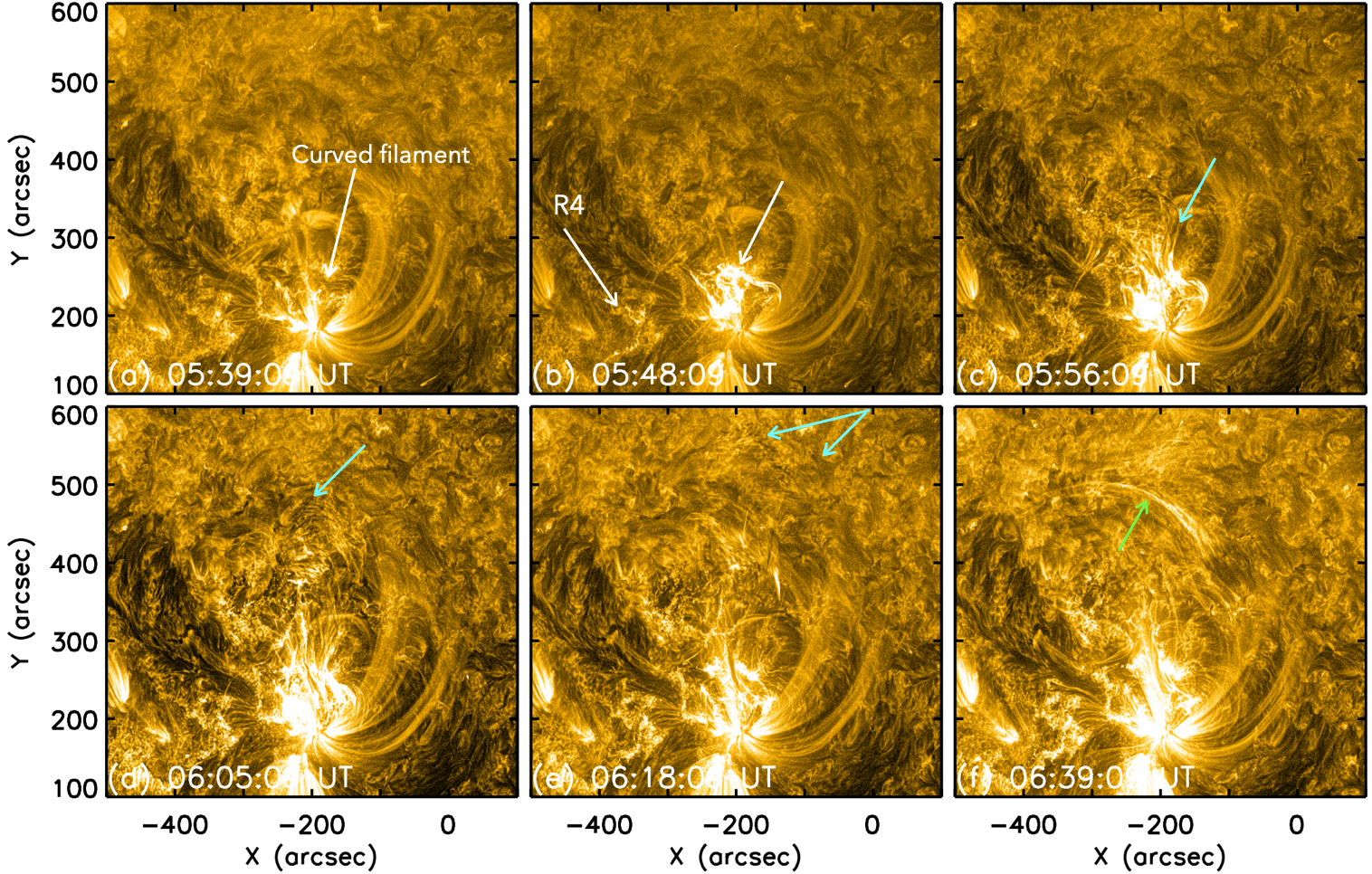}
\caption{Evolution of the eruption in AIA 171 stressing mainly its later stages. Part of the curved filament before eruption is pointed with a {\it white arrow} in panel a. The elongated ribbon R4, as well as the up-going material, appear in panel b as indicated by the {\it white arrows}. Panels c, d, and e show parts of the cool material going upward embedded in hot plasma (see the {\it cyan arrows}). Plasma falling back towards the solar surface after reaching its maximum height is indicated with a {\it green arrow} in panel f. The data are processed using the MGN technique for the better visibility. See the accompanying movie \url{AIAMGN171_09May2019_Fig9.mp4}.}
\label{evolution171}
\end{figure*}

\subsubsection{The Failed Eruption from Earth's Point of View}
\label{sec:failed_AIA}

In this section we discuss the different stages of the filament eruption as seen from Earth. We refer to the previously described Figs.~\ref{AIA} and \ref{IRIS}, stressing the aspects relevant to the eruption (see also movies \url{AIA304_09May2019_Fig3_Fig6.mp4} and \url{IRIS_CII_09May2019_Fig7.mp4}). To these figures, we add Figs.~\ref{Halphaevolution} and~\ref{evolution171} that depict a larger FOV. 

In \sect{flare} we have shown the existence of the two ribbons related to the C6.7 flare located in the center of the active region, respectively, R1+2R1 and R2+2R2. The later is the L-shape ribbon  well visible in Fig.~\ref{IRIS}d at 05:47 UT. Before this time, we already observe the lift off of part 1 of the filament as indicated by the black arrow in Fig.~\ref{IRIS}b.  In the movie of IRIS we observe that this  brightening becomes more diffuse and extends.  Around 05:47 UT part 2 of the filament escapes.    
 To the north of the two ribbons, we clearly see a  large diffuse area  with a  bright northern edge oriented NE-SW (black arrow in Fig.~\ref{IRIS}d). This large diffuse area is also visible later on at 05:49 UT in AIA 304 \AA\  (Fig.~\ref{AIA}g). 
As part 2 lifts, the whole filament appears as a large flux rope with a NE-SW orientation after 05:49 UT (black arrows in Fig.~\ref{AIA}g, k, h and l).

The evolution just described, as well as the location of the two main ribbons, described in \sect{flare}, lets us speculate that probably magnetic flux cancellation at sites \texttt{b} and \texttt{c} (see Figs.~\ref{twoCRs} and~\ref{MMF} and \sect{obs_B}) may have played a role in the filament destabilization and eruption. In the higher temperature AIA band, AIA 171, the most evident feature is the presence of the heated plasma extending upward in Fig.~\ref{AIA}j,k,l; notice that part of the filament plasma seems to be flowing back already at around 05:51 UT.

We can continue observing
the journey of the erupting plasma in H$\alpha$ and AIA 171  in a larger FOV 
in Figs.~\ref{Halphaevolution} and~\ref{evolution171}, and the corresponding movies after 06:00 UT until $\approx$ 06:20 UT.  
The  AR  viewed in AIA 171 is covered by a bright area of loops and straight features to its north (see white and green arrows in Fig.~\ref{evolution171}b,c). 
The H$\alpha$ material is seen to move upwards in Fig.~\ref{Halphaevolution}d,e.  However, simultaneously, the plasma  is also observed falling down, dark in H$\alpha$  and bright in AIA 171 in both panels f of Fig.~\ref{Halphaevolution} and Fig.~\ref{evolution171}. The falling down material is  progressively stack along large scale loops mostly visible in AIA 171 until 06:39 UT (Fig~\ref{evolution171}f).

\begin{figure*}[t!]
\centering
\includegraphics[width=\textwidth]{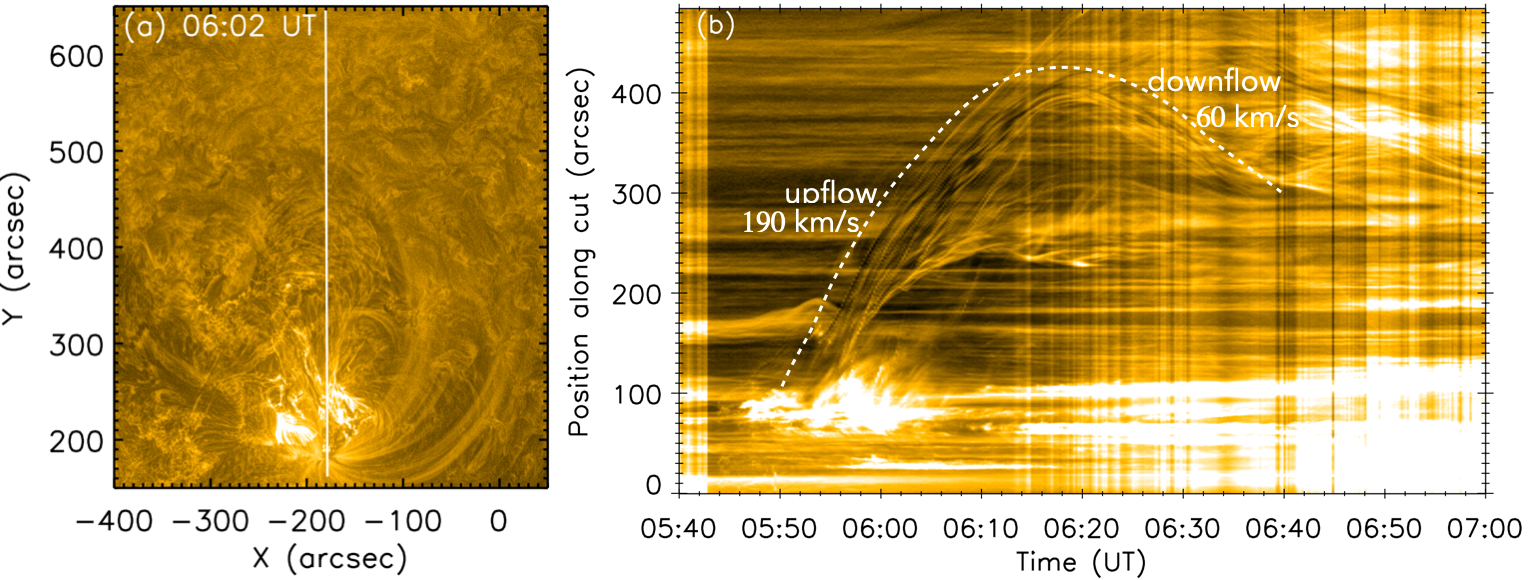}
\caption{Height-time analysis of the filament eruption in AIA 171. The upward and downward projected motion of the plasma is measured along the N-S white slit in panel a on the AIA 171 image. The slit for constructing the stack plot is chosen manually by eye and roughly agrees with the central location of the filament part 1. Panel b corresponds to a stack plot built along the slit. For a better visibility of the upward- and downward-moving material, we have used the MGN technique to process the images used to build this plot. The {\it white-dashed line} in this panel is drawn manually, following the leading edge of the material moving first upward and later downward.}
\label{timeslice}
\end{figure*}

To evaluate the speed of the rising and falling plasma we have built a stack plot along the N-S line that is shown in Fig.~\ref{timeslice}a. Since the eruption of the filament, as well as the falling back of the plasma, is complex and appears to occur at different stages and along different directions, we have chosen only one direction that roughly agrees with the central location of the filament part 1 to have average speed estimations. Using the slopes of the white-dashed line, drawn by hand in Fig.~\ref{timeslice}b, which follows the leading edge of the material along the N-S line in panel a, we estimate a speed projected on the plane-of-the-sky of 190 km s$^{-1}$ for the upflow and 60 km s$^{-1}$  for the downflow. 
We deproject these values in the direction of the eruption using the angles measured in EUVI-A 304 and 195 in \sect{failed_STEREO}; when doing so, we obtain 201 km s$^{-1}$ and 62 km s$^{-1}$, respectively.  These values are quite similar to those found on the plane-of-the-sky because of the very small angle between the directions N-S and that of the eruption. 
We also measure the distance reached by the plasma along the N-S direction, computed from around 70$\arcsec$ in Fig.~\ref{timeslice}b where the intense flare emission is seen in AIA 171, and find a value of $\approx 270\,\mathrm{Mm}$; this corresponds to a distance of $\approx280\,\mathrm{Mm}$ along the plasma ejection direction (assumed to be along a straight line).

\section{Coronal Field Model of the Events on 9 May 2019}
\label{sec:model}

\subsection{Overview}
\label{sec:overview}

Following our multi-wavelength analysis of the phenomena in AR 12740 on 9 May 2019, we present in \sect{local_model}  a flare model of the magnetic field configuration at the AR scale size. The field line connectivity derived from this model lets us propose a possible physical scenario (\sect{scenario}) and interpretation of the complex chain of events we have analysed. This section is followed by a global magnetic field model  (\sect{global_model}) that complements and supports our proposed scenario and interpretation.

\subsection{Local Magnetic Field Model}
\label{sec:local_model}

To understand the role of the different magnetic polarities in AR 12740, we model its coronal field. We extrapolate the HMI LOS magnetic field to the corona using the discrete fast Fourier transform method described by \citet{Alissandrakis1981}, under the linear force-free field (LFFF) approach ($\curl \vec B = \alpha \vec B$, with $\alpha$ constant). Although this kind of modeling cannot take into account the distribution of currents at the photospheric level and the strong shear that we can infer from the shape and location of the ribbons of the two-ribbon flare, but only the shear in the global magnetic configuration, its computation is fast and has proven to be efficient to determine the magnetic field structure at the scale size of an AR, which can be later compared with observed 
active events \citep[see e.g.][and references therein]{Mandrini2006,Mandrini2014}. 

\begin{figure*}[t!]
\centering
\includegraphics[width=\textwidth]{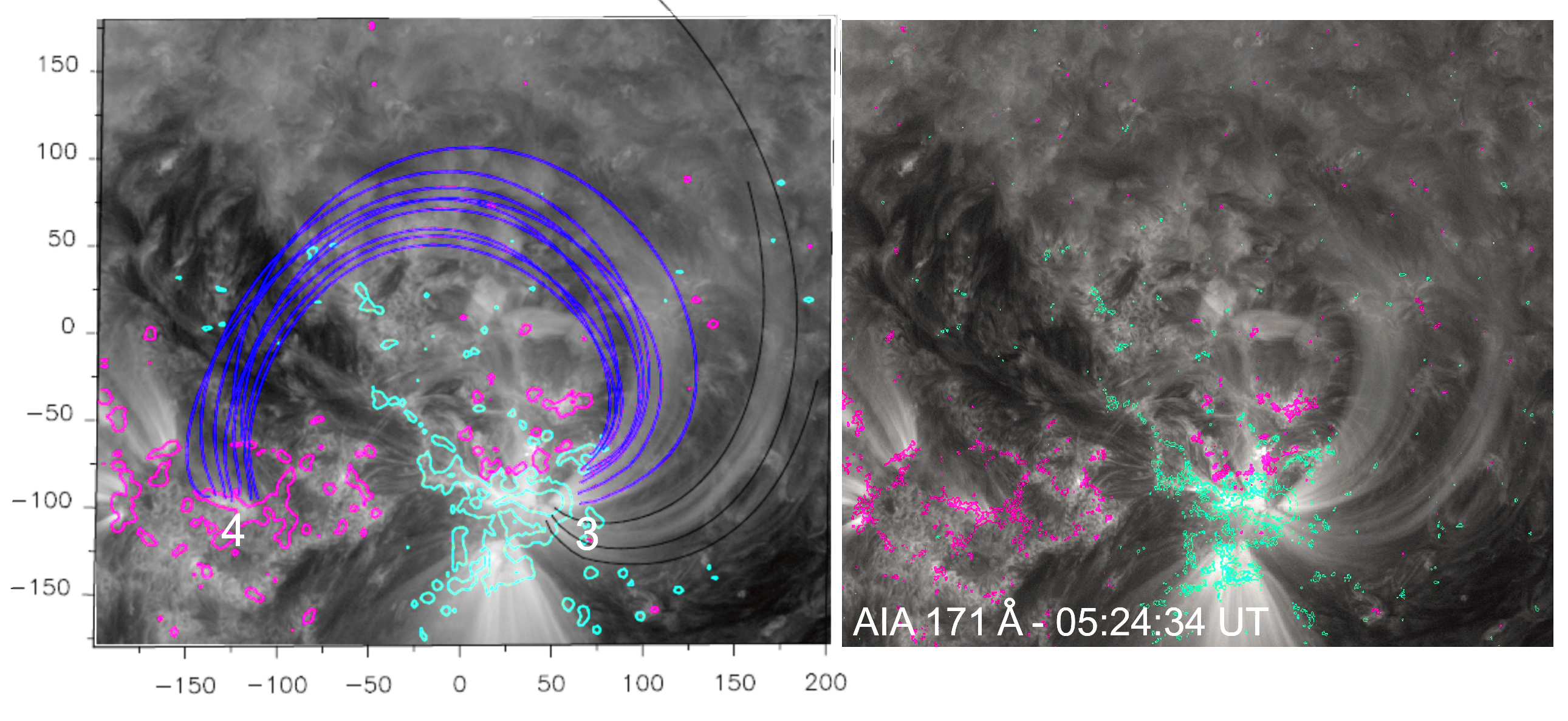}
\caption{{\it Left panel}: Magnetic field model of the large scale coronal loops connecting the chain of small negative polarities labeled as 3 to the disperse positive following AR polarity 4 (see Fig.~\ref{twoCRs}). A set of computed field lines in {\it blue solid traces} is overlaid on the AIA 171 image at 05:24:34 UT, together with HMI magnetic field contours ($\pm$ 100, 500 G, positive
(negative) shown in {\it magenta (blue)} color). The three {\it black field lines} connect to northern positive quiet-Sun regions out of the AR (compare to Fig.~\ref{AIA-PFSS}). The axes in this panel are in Mm, with the origin set at the AR center. {\it Right panel}: The same AIA image shown as background in the left panel for comparison. The image is shown in logarithmic direct intensity and we have added HMI isocontours of similar values to those in the left panel as a reference. The images are shown in {\it grey scale} to facilitate the visualization of computed field lines and magnetic field contours in this figure and the following two.} 
\label{preflare}
\end{figure*}

\begin{figure*}[ht!]
\centering
\includegraphics[width=\textwidth]{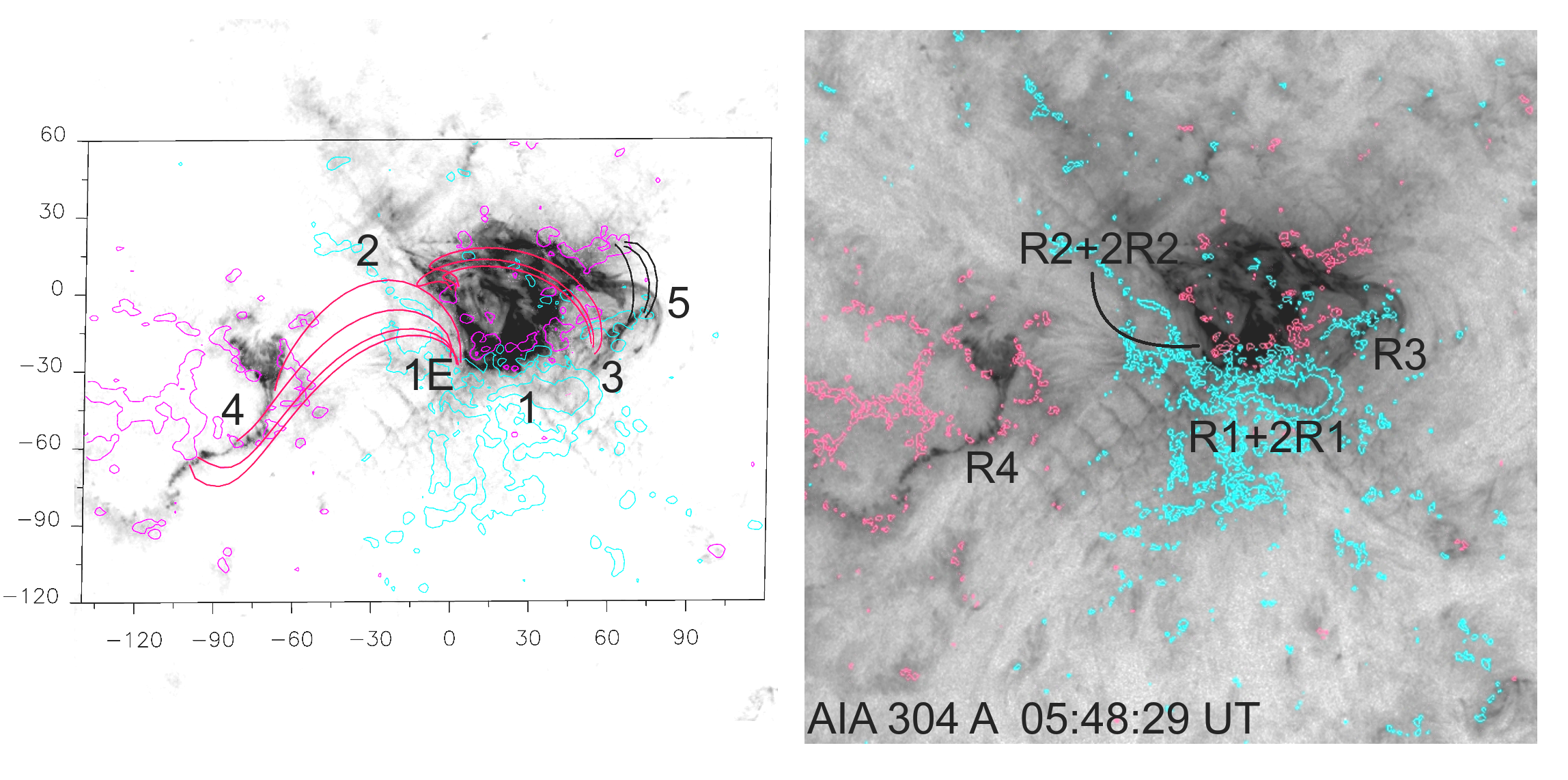}
\caption{{\it Left panel}: Magnetic field model at the flare time. Sets of field lines in {\it continuous tracings} are overlaid on the AIA 304 image at 05:48:29 UT. The set in {\it red} color to the east (west) connects the ribbon on polarity 4 (3) to the one on polarity 1 (2) and is the result of the external reconnection process discussed in the text (see \sect{scenario}). The set in {\it black} color has been added to show that the negative polarity 5 (see Fig~\ref{twoCRs}), where a curved brightening to the north of R3 is located, is connected to a northern positive polarity. The convention for HMI contours and axes are the same as in Fig.~\ref{preflare}. {\it Right panel}: The same AIA image shown as background in the left panel in logarithmic reverse intensity including HMI contours for reference, notice the diffraction pattern because of the high flare intensity. The two-ribbon flare, 2R1 and 2R2 on polarities 1 and 2, and the ribbons of the quadrupolar external reconnection have been labelled in this panel. R1 and R2 are located on polarities 1 and 2 and, as discussed in the text, they cannot be clearly separated from the two main flare ribbons. Ribbon R3 is located on the chain of small negative polarities 3 and the extended ribbon R4 is located on 4.}
\label{flare}
\end{figure*}

\begin{figure*}[ht!]
\centering
\includegraphics[width=\textwidth]{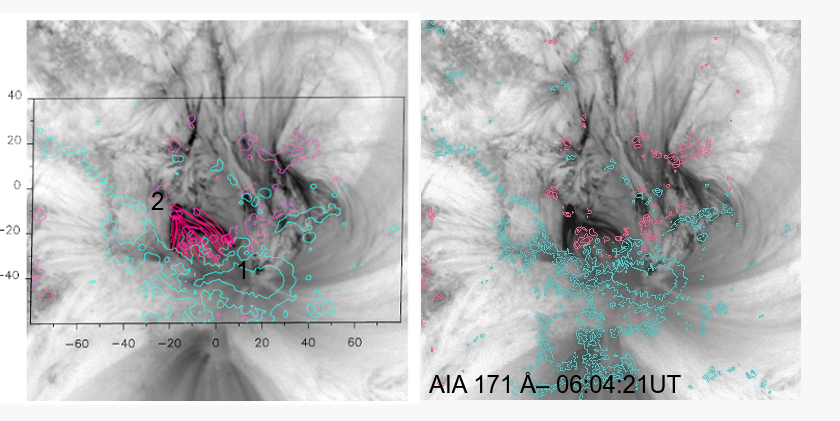}
\caption{{\it Left panel}: Magnetic field model during the flare decay phase. A set of computed field lines in {\it red continuous tracing red solid traces} is overlaid on the AIA 171 image at 06:04:21 UT. This set corresponds to the loops connecting the two flare main ribbons and results from the internal reconnection process discussed in the text. The field lines are anchored to polarities 1 and 2. The conventions for HMI contours and axes are the same as in Fig.~\ref{preflare}. {\it Right panel}: The same AIA image shown as background in the left panel for comparison including HMI contours and using the same convention as in Fig.~\ref{flare}. Notice that some loops can be discerned between polarity 3 and a northern positive polarity as expected from the {\it black lines} added to Fig.~\ref{flare}.}
\label{posflare}
\end{figure*}

Figure~\ref{preflare} right panel shows an AIA 171 image before the flare (05:24:34 UT) in which large scale magnetic loops are visible. Figure~\ref{preflare} left panel  displays a set of blue field lines derived from the coronal model overlaid on the same AIA image. In this and all other coronal field models, we use as boundary condition, the HMI magnetogram closest in time and we also apply a transformation of coordinates from the local frame, in which the computations are done, to the observed frame so that our models can be directly compared to the data \citep[see the Appendix in][]{Demoulin1997}. The value of $\alpha$, the free parameter of the model, is set to best match these large-scale loops \citep[as discussed in][]{Green2002}. The best-matching value is $\alpha$ = 9.4 $\times$ 10$^{-3}$ Mm$^{-1}$. This large-scale loops are also present several hours after the flare has ended, as can be seen in images displayed in Helioviewer  (\url{helioviewer.org/}), which means that the large-scale configuration persists.

Figure~\ref{flare} right panel depicts an AIA 304 image 3 min before flare maximum (05:48:29 UT). The flare ribbons corresponding to the two-ribbon flare as those associated to the quadrupolar configuration have been labelled as indicated in \sect{flare}. Figure~\ref{flare} left panel displays a set of red field lines derived from the coronal model overlaid on the same AIA image.  Since no flare loop is observed to compare with our computed field lines, the value of $\alpha$ is set so that the computed field lines connect the observed ribbons. The best-connecting value is higher than in Fig.~\ref{preflare}, $\alpha$ = 1.6 $\times$ 10$^{-2}$ Mm$^{-1}$ and double this value for the sets of lines to the East and West, respectively. See the caption to this figure for an explanation of the set of black lines. 

Finally, we model the loops observed during the flare decay phase that are observed between the two main flare ribbons. Figure~\ref{posflare} right panel depicts an AIA 171 image at 06:04:21 UT, where the so-called post-flare loops are clearly seen. Figure~\ref{posflare} left panel displays a set of red field lines overlaid on the same AIA image that match the shape of these post-flare loops. The value of $\alpha$ that gives the best match is $\alpha$ = 1.6 $\times$ 10$^{-2}$ Mm$^{-1}$.

\subsection{The Stages of the Observed Events}
\label{sec:scenario}

\begin{figure*}[ht!]
\centering
\includegraphics[width=\textwidth]{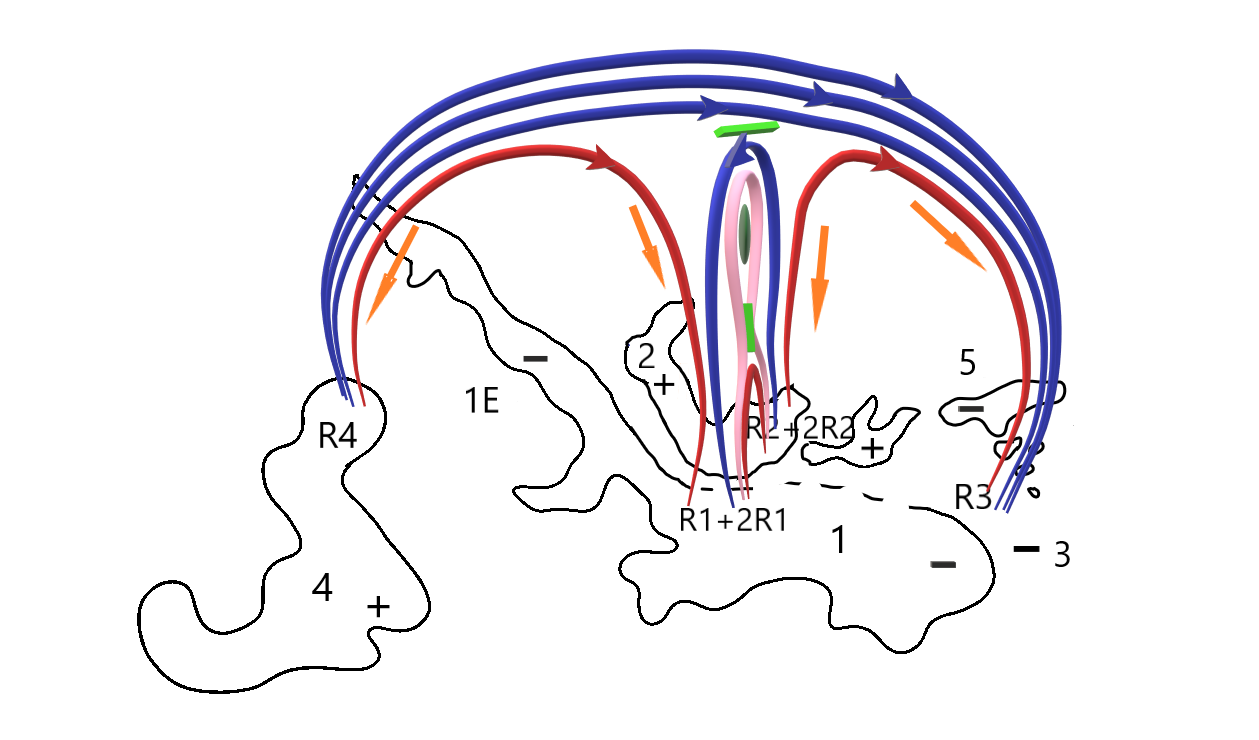}
\caption{Sketch showing sets of field lines connecting several of the 
polarities identified in Fig.~\ref{twoCRs}. The relative locations and shapes of these polarities have been drawn and the sites of the different ribbons are indicated using the labels in Fig.~\ref{flare}. The flux rope configuration including the filament is simplified to a 2D representation. As the filament (indicated as a {\it green ellipse}) and its magnetic configuration 
rise two reconnection processes occur, as identified with {\it thick green segments}: the internal one below the filament and the external one above it. The former process gives the observed two main flare ribbons (2R1 and 2R2) located close to the base of the {\it pink line} and joined by a {\it red} reconnected field line. The latter process reconnects blue field lines (connecting regions on polarities 3 and 4) with {\it blue} elongated field lines (connecting 1 and 2), which overlay the rising filament. This process eventually derives in the injection of filament material in field lines connecting 1 to 4 and 3 to 2, highlighted in red color. This material is observed flowing down (see {\it orange arrows}) along them pinpointing the filament failed eruption (see \sect{scenario} for a more detailed description). 
At the footpoints of the {\it red lines} we observe ribbons R1 and R2, which cannot be clearly separated from the two main flare ribbons resulting from the internal reconnection process, and the farther ribbons R3 and R4.}
\label{sketch}
\end{figure*}

The results of the three just described models, together with our data analysis, leads us to the following conclusions about the origin of the C6.7 flare and its evolution.
The connectivity determined from each model provides only a static view at the time it is computed; therefore, to facilitate our discussion of the different stages of the events and the processes that occur, we include the scheme shown in Fig.~\ref{sketch}. This sketch is similar to the one proposed by \citet{Lopez-Fuentes2018} and \citet{Poisson2020} for a failed mini-filament eruption.

The situation depicted in Fig.~\ref{sketch} corresponds to a time at which the magnetic configuration 
containing the filament, drawn as a green oval, was already destabilized and rising. Probably, magnetic flux cancellation occurring at sites \texttt{b} and \texttt{c}  destabilizes the filament magnetic configuration 
that starts erupting (see references in \sect{intro} about filament eruptions driven by flux cancellation). A set of field lines (anchored between 1--2) overlays the filament that is located along the main AR PIL (see the elongated blue line lying above the filament). The set of long blue field lines (anchored between 3 and 4, see Fig~\ref{preflare} left panel and Fig.~\ref{sketch}) corresponds to the closed background field. 

Magnetic reconnection sets below the filament as happens in a classical prominence eruption \citep[see e.g. ][]{Aulanier2010,Webb2012}. The location of this reconnection process is represented by the green vertical segment in the sketch. The pink field line marks the limit between the reconnected field lines below the filament and those surrounding it. In our model, the red field lines in Fig.~\ref{posflare} left panel correspond to the reconnected lines resulting from this process below this limiting pink line. The just described reconnection process has been called internal in several articles \citep[see e.g.][]{Sterling2015, Moore2018}. 

As the filament configuration moves up, field lines located above the filament (the blue elongated line in the sketch) start reconnecting with the large-scale blue lines shown in our model in Fig.~\ref{preflare} left panel and outlined in blue in the sketch. This second reconnection process, indicated by the green oblique segment, has been called external in the just mentioned references.

As a result of the external reconnection process, the filament plasma and that of the loops where it is still embedded, is injected into the red reconnected field lines. The material is seen flowing down along them (as indicated by the orange arrows in the sketch) and the eruption fails. The external reconnection process decreases the magnetic tension above the filament flux rope. However, if the large-scale magnetic field (in the blue arcade connecting 3 to 4) has more flux than that of the flux rope, the later could be mostly reconnected and could not continue upward. 

To investigate the latter statement, we first compute the magnetic flux swept by the ribbons of the two-ribbon flare using AIA 1600 images overlaid on the corresponding HMI magnetograms (the two ribbons are better seen and not saturated in this AIA band). The flux swept by the ribbons represents the flux added by reconnection to the erupting flux rope \citep[see e.g. ][and references therein]{Deng2017} and is a lower bound for the flux rope total flux. This estimated average flux is $\approx$ 1.5$\times$10$^{20}$ Mx for the time range 05:43\,--\,05:53 UT (see AIA 1600 images in Fig.~\ref{AIA}a-d).  As a second step, we compute the flux in the large-scale overlying arcade taking into account the region  on polarity 4 that connects to polarity 3 in our local field model (we use only polarity 4 because the counterpart region on polarity 3 is continuously evolving because of the shuffling of MMFs).  The flux in the large scale arcade is $\approx$ 8.5$\times$10$^{20}$ Mx, $\approx$ 6 times larger than the flux rope flux. This supports our assumption of a fully reconnected erupting flux rope. In addition to this, the kinetic energy of the filament with a speed of 183 km s$^{-1}$ could be too small for a successful eruption, see e.g. \citet{Shen2011} who studied three filament eruptions, two failed and one successful, and found that the filament velocity in the successful one was the largest and that filament velocities were proportional to the power of their flares.

In summary, the first internal reconnection process would result in the observed intense two ribbons labelled as 2R1 and 2R2 in Figs.~\ref{flare} (right panel) and~\ref{sketch}. They are located at both sides of the PIL between polarities 1 and 2, and the very short post-flare loops joining them (see Fig.~\ref{posflare} left panel). The second external reconnection process is associated to ribbons R4, R3 and their counterparts on polarities 1 and 2 that we have called R1 and R2 (Figs.~\ref{flare} right panel and~\ref{sketch}). As already mentioned, these ribbons cannot be clearly separated from the main two flare ribbons and we just only have an idea of their location based on the field-line connectivity computed from our model shown in Fig.~\ref{flare} left panel. 
In the studied event, as stated above, the second external reconnection process is most relevant in impeding the filament flux rope eruption.

\subsection{Global Magnetic Field Model}
\label{sec:global_model}

Since our local field model is limited to the scale size of the AR, we have computed a global coronal magnetic model to verify that the magnetic configuration at this larger scale remains closed.  

\begin{figure*}[t!]
\centering
\includegraphics[width=\textwidth]{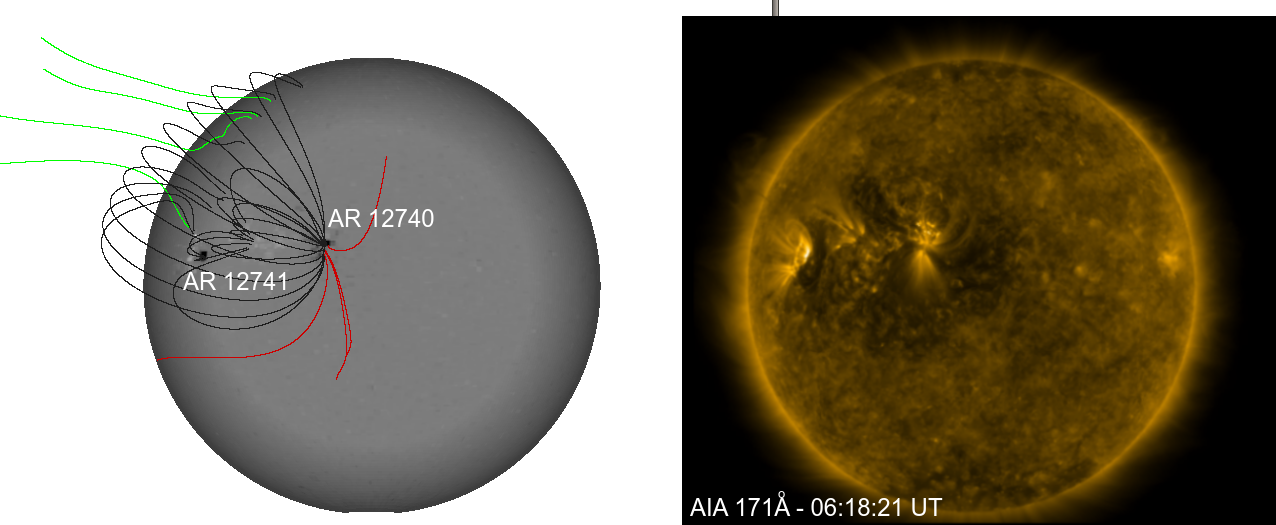}
\caption{{\it Left panel}: PFSS model of CR 2217 with AR 12740 located at Carrington longitude 318$\degree$ on 9 May 2019 close to the flare time. The field-line color convention is such that {\it black} indicates closed lines and {\it pink (green)} corresponds to open lines belonging to the negative polarity  (positive polarity) field (notice that open implies reaching the source surface). Closed field lines connect the AR main negative sunspot to its following positive polarity as in the model in Fig.~\ref{preflare}. Notice that the curvature of this set of closed field lines is different from the set shown in {\it blue} in that figure since this is a potential field model, while the local field model considers the shear at the AR scale-size. Other closed lines connect north to quiet-Sun regions as in  Fig.~\ref{preflare} (lines shown in {\it black}) or to the positive field of the trailing AR 12741. The magnetic field values have been smoothed and saturated above (below) 250 G (-250 G). {\it Right panel}: AIA 171 full disk image as reference at the time corresponding to the Carrington longitude in the left panel.}
\label{AIA-PFSS}
\end{figure*}

\begin{figure*}[t!]
\centering
\includegraphics[width=11.cm]{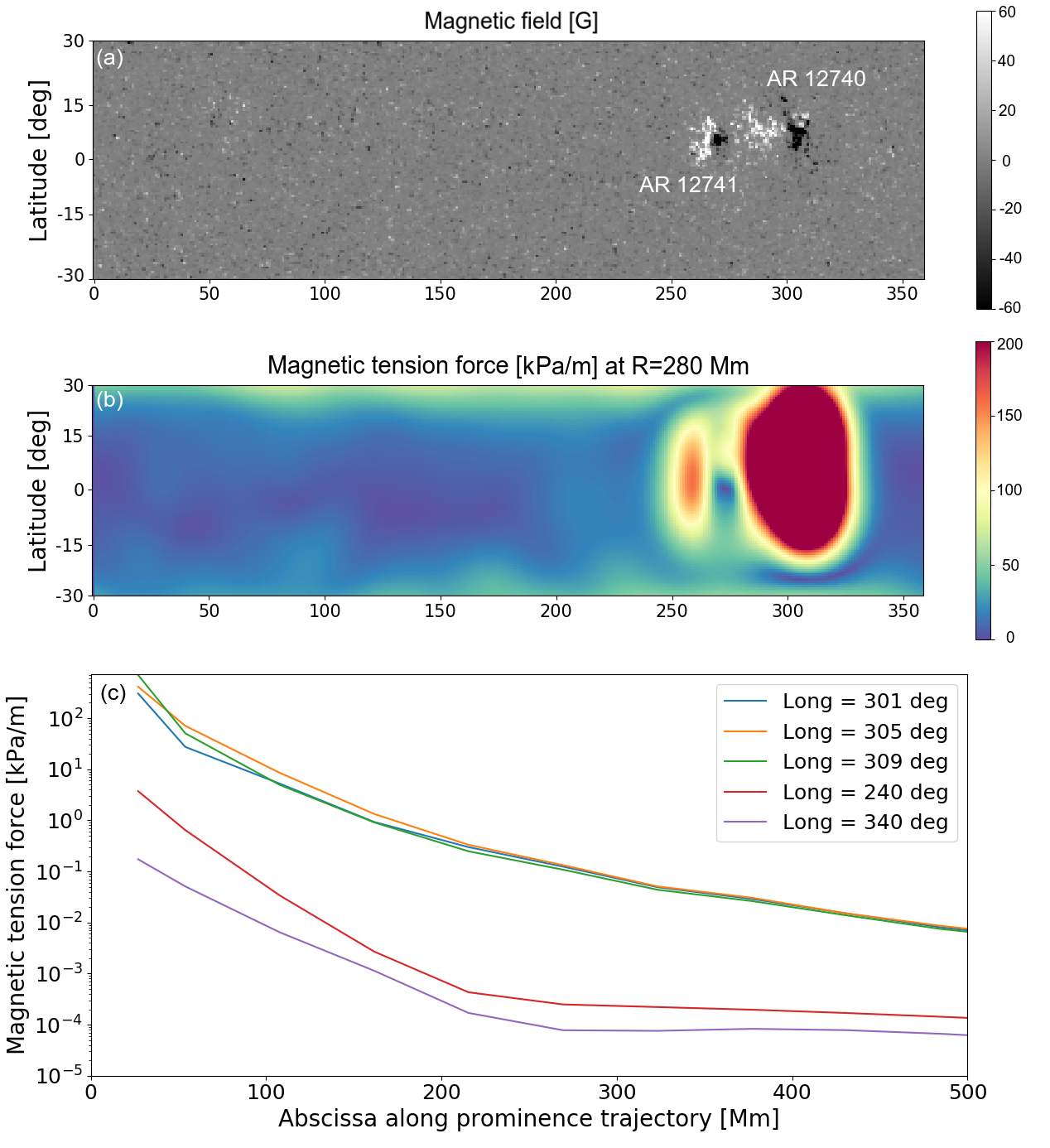}
\caption{From top to bottom: (a) Synoptic HMI map of CR 2217.  The horizontal axis indicates the Carrington longitude and the vertical axis on the left corresponds to the Carrington latitude.  Note that we have limited the latitudinal extension to $\pm$ 35 $\deg$ to exclude field lines that are considered open in the PFSS model. A  {\it greyscale bar} showing the magnetic field scale intensity has been added to the right. 
(b) Magnetic tension force at a height of 280 Mm, with the same coordinates than in (a).
A {\it color bar} showing the magnetic tension scale has been added to the right. 
(c) Magnetic tension force in function of the coordinate along the prominence trajectory computed at three different longitude values at the AR location. The trajectory is assumed to be a straight line, inclined to the local vertical as observed (Fig.~\ref{fig:stereo}) and set within a fixed meridian plane. We have also added two curves computed in a similar way at two different longitudes to the east and west of ARs 12741 and 12740  for comparison. } 
\label{tension}
\end{figure*}

The  global  coronal  magnetic  field  of  CR 2217  is  modeled using a potential-field source-surface (PFSS) approach. These models assume a current-free coronal field with an observationally prescribed boundary condition at the photosphere. PFSS models  assume  that  the field becomes purely radial  at  a  given  height  called  the source  surface, which in our case is set to a value of 2.5 \Rsun. Our PFSS model uses as its lower boundary condition the corresponding HMI magnetic field synoptic map. 

The model is done using the Finite Difference Iterative  Potential-Field  Solver  (FDIPS) code described  by  \citet{Toth2011}. This  code is  freely  available  from  the  Center  for  Space Environment  Modeling  (CSEM)  at  the  University  of Michigan  (\url{csem.engin.umich.edu/tools/FDIPS}). It uses an  iterative  finite-difference  method  to solve the Laplace equation for the magnetic field. In this particular case, the spatial resolution is 
$1\degree$ in longitude (360 longitudinal grid points), 0.011 in the sine of latitude (180 latitudinal grid points) and 0.01 \Rsun\ in the radial  direction. 

Figure~\ref{AIA-PFSS} left panel shows the result of our modeling together with a set of field lines computed starting integration at a height of 150 Mm in both directions. The integration points are located in AR 12740, its neighborhood, and the trailing AR 12741. It is clear that the magnetic field configuration remains closed at the large scale, with closed lines connecting the leading and following AR polarities and also the leading negative polarity to quiet-Sun regions located far to the north of the AR. Figure~\ref{AIA-PFSS} right panel shows an AIA 171 full disk image as reference, the large scale loops connecting both AR 12740 polarities are clearly seen. 

We have computed the magnetic tension force or magnetic tension,
$\vec B \cdot \nabla \vec B/\mu_0$, with $\vec B$ the three components of the
magnetic field directly derived from the PFSS model and $\mu_0$ the vacuum magnetic permeability), at different heights (Fig.~\ref{tension}). 
The magnetic tension is directed towards the centre of curvature of the field lines and acts as a restoring force which works against the ejected magnetic field. 
Figure~\ref{tension}a shows the HMI synoptic map for CR 2217 which helps us identify the locations of AR 12740 and the trailing AR 12741. 
Figure~\ref{tension}b shows that the magnetic tension force is the largest over both ARs compared to the surrounding. This is so over a broad interval of heights (at least up to 500 Mm), while the prominence stays confined lower down (Fig.~\ref{fig:stereo}). We also compute the magnetic tension force along the prominence trajectory, approximated by a straight line inclined to the local vertical as observed by STEREO A (Fig.~\ref{fig:stereo}). Figure~\ref{tension}c shows the results for the trajectory located within three meridional planes.  While the magnetic tension decreases fast along the trajectory, it still stays large compared to the surroundings (panel b). Notice that we have also added two curves computed in a similar way, but at both sides ({\it east and west}) of ARs 12741 and 12740, to stress the difference between the values of the tension in the surroundings to those in the AR where our events occurred. 
This lets us conclude that it is the magnetic tension of the overlaying field that prevents the filament configuration to erupt (see references in \sect{intro}). Despite the fact that the magnetic tension can be decreased by forced reconnection between the erupting magnetic field and the overlying arcade, the later has enough magnetic flux and intensity to stop the filament eruption at a moderate height.  

\section{Summary and Conclusions}
\label{sec:discussion}
We analyse a series of events that occurred in AR 12740 on 9 May 2019 using a set of multiwavelength observations going from the photosphere to the corona obtained by HMI, AIA, STEREO, IRIS, and GONG/LSO instruments. The chain of events includes the formation of a filament, its destabilization and the accompanying flare, followed by the filament failed eruption. Our study lets us conclude on the origin of each of the different steps in this chain. 

AR 12740 was in its decaying phase characterized by the presence of MMFs surrounding a compact and high-intensity field negative leading polarity followed by a very disperse positive one. Though the AR could be globally considered as bipolar, the constant advection of minor polarities from the main spot into the surrounding moat region and the emergence of small bipoles created a very complex and dynamic magnetic configuration. A detailed study of the magnetic field evolution leads us to identify four main polarities that played a key role during the filament eruption and flare, i.e. these two events occurred within a mainly quadrupolar AR (see \sect{obs_B}). Magnetic flux cancellation within the moat region to the north of the main spot, in a site that we called \texttt{a} (see \sects{obs_B}{filform}), was the origin of the formation of a long and curved filament by reconnection  between sets of fibrils, like in the model proposed by \citet{vanBallegooijen1989} \citep[see an observed example in][]{Schmieder2004}.

In a similar manner, magnetic flux cancellation was at the origin of the destabilization of the flux rope containing the filament plasma (see \sects{obs_B}{filerup}). This mechanism  was proposed in several examples and simulations of minifilament eruptions followed by blow-out jets (see reference in \sect{intro}). The flux cancellation process was mainly due to the constant shuffling of the MMFs at two different sites (sites \texttt{b} and \texttt{c}) by the PIL around the main spot.  This eruption was accompanied by a two-ribbon flare whose main ribbons were located on the main negative polarity and an L-shape positive polarity to its NE (see \sect{flare}). However, because the global magnetic configuration of AR 12740 was quadrupolar, two additional ribbons were seen far to the {\it east and west} of the two-ribbon flare. A force-free magnetic field model at the AR scale size allows us connecting the far flare ribbons between themselves and to the extensions of the two main flare ribbons, {\it i.e.} the flare was in fact a six-ribbon event confined by the larger scale loops of the quadrupolar configuration (see \sects{local_model}{scenario}). 

Even though the flux rope containing the filament erupted, this eruption failed thus plasma was observed first moving upwards and later downwards. Based on our local magnetic field model, we propose a scenario (see \sect{scenario}) in which the failed eruption and multi-ribbon flare are the result of two reconnection processes, one occurring below the erupting flux rope, leading to the two-ribbon flare, and another one above it between the filament configuration and the large-scale closed loops of the quadrupolar configuration. 
This second process leads to the appearance of the far flare ribbons and their counterparts (as extensions of the main two ribbons). In a similar way, it injects plasma from the filament and the loops where it is embedded, within the reconnected loops linking the ribbons of the quadrupolar configuration. These two reconnection processes have been called internal and external in articles describing minifilament eruptions \citep[see e.g.][and references therein]{Sterling2015, Moore2018}.  
Furthermore via this external reconnection process, the erupting flux rope could fully reconnect with the large-scale closed loops because, as we have shown, its magnetic flux is much lower.    
A PFSS model confirms that AR 12740 was confined by closed field lines connecting both AR main polarities and the main negative polarity to quiet-Sun regions. Additionally, from this model we compute the magnetic tension of the large-scale magnetic field at a height above that one reached by the erupting plasma and conclude that above the AR it was much larger than in other locations on the Sun (see \sect{global_model}). Therefore, from the point of view of the global magnetic configuration, we also find hints that  would lead to a failed filament eruption.

Summarizing, from an observational point of view, this case study is clearly consistent with models proposing that filaments can be formed from converging fibrils at flux cancellations sites, as well as destabilized by similar flux cancellation processes (see references in \sect{intro}).
Furthermore, it represents a well observed example of how magnetic confinement by an intense overlying field can lead to failed flux rope eruptions, as proposed by several MHD simulations \citep[e.g.][]{Fan2003,Amari2018}.

\appendix
\section*{Study of the CME visible by STEREO-A COR1}
\label{S-appendix}

For completeness, we analyse a CME whose leading edge appears in the STEREO-A COR1 FOV at 05:55 UT, which we briefly describe here. We also refer the reader to the movies that can be generated at \url{cdaw.gsfc.nasa.gov/stereo/daily_movies/} for a quick look of the CME event. Its projected speed at the central position angle as measured in COR1-A images yields 246 km s$^{-1}$. When reaching the COR2-A FOV, the CME appears faint and diffuse, but is nonetheless detected by the Solar Eruption Detection System (SEEDS, \url{spaceweather.gmu.edu/seeds/monthly.php?a=2019&b=05&cor2}) at George Mason University. In this data base, the CME position angle is $260\degree$,  {\it i.e.} $10\degree$ south from the solar equator, and its speed in the plane-of-the-sky is 233 km s$^{-1}$, in good agreement with the value we compute from COR1-A images.

Although it is out of the scope of our work, we attempted --unsuccessfully-- to identify the CME source region in AIA images, as well as in the H$\alpha$ ones from LSO. Further inspection, now from the quadrature vantage point provided by STEREO-A, shows coronal material at a fairly high altitude ($\sim$\,1.4\,\Rsun) and at a position angle of $\approx285\degree$, that starts moving outward in a radial fashion, apparently destabilized and triggered by the flare we study. Given this scenario, we speculate that we are dealing with a stealth event, originating due to the destabilization of a barely visible structure which lies at a significant height above the solar surface, already prior to the start of the C6.7 flare.
Being this CME eruption magnetically connected to the failed eruption or not, both can be regarded as separate events due to a number of reasons. Firstly, and as mentioned above and shown in Fig.~\ref{fig:stereo}, the failed eruption moves with an angle of $\approx 60\degree$ with respect to the radial direction, whilst the outward-travelling coronal material seen at a high altitude propagates nearly radially. Secondly, the failed filament eruption is seen to turn back at $\sim$06:30 UT, while the CME at that time is at $\approx$\,3\,\Rsun~and reaches the COR2-A FOV at 06:54 UT. Therefore, because of timing and propagation direction, compared to that of the filament failed eruption seen in EUVI-A images, we conclude that the CME observed in COR1-A and COR2-A can be regarded as not affecting the events analysed in this article.

%
 \begin{acks}
We thank the reviewer for his/her useful comments and suggestions.
We thank the open data policy of SDO, GONG, IRIS, 
STEREO instruments. IRIS is a NASA small explorer mission developed and operated by LMSAL with mission operations executed at NASA Ames Research Center and major contributions to downlink communications funded by ESA and the Norwegian Space Centre. This work was initiated by R. Joshi, B. Schmieder, and P. Démoulin at the Observatoire de Paris, Meudon. C.H. Mandrini thanks the Observatoire de Paris, Meudon for an invitation. We made use of NASA's Astrophysics Data System Bibliographic Services.
We recognise the collaborative and open nature of knowledge creation and dissemination, under the control of the academic community as expressed by Camille No\^{u}s at www.cogitamus.fr/indexen.html.
\end{acks}

\begin{authorcontribution}
RJ did the data analysis and wrote the draft of the paper. CHM, RC, and BS wrote the substantial parts of the manuscript and contributed to the interpretation. CHM did the local magnetic field modeling and CMC did the global one. GDC contributed to analysis of the magnetic field observations and related computations. HC contributed to the analysis of the coronal data and the CME observations. PD helped with the physical interpretation of the observations. All the authors did a careful proofreading of the text and references.
\end{authorcontribution}

\begin{funding}
This research is supported by the Research Council of Norway through its Centres of Excellence scheme, project number 262622. RJ thanks to Indo-French Centre for the Promotion of Advanced Research for a Raman Charpak Fellowship. RC acknowledge the support from Indo-Bulgarian bilateral project by Department and Science and Technology, New Delhi, India. CHM, GDC, HC and CMC acknowledge grants PICT 2016-0221 (ANPCyT) and UBACyT 20020170100611BA. HC and CHM appreciate support from grant MSTCAME8181TC (UTN) and HC from PIP 11220200102710CO (CONICET). GDC and HC are members of the Carrera del Investigador Cient\'\i fico of the Consejo Nacional de Investigaciones Cient\'\i ficas y T\'ecnicas (CONICET). CHM is a CONICET researcher and CMC is a CONICET fellow. 
\end{funding}

\begin{dat}
The datasets analysed during the current study are available at  \url{https://iris.lmsal.com/data.html}, \url{http://jsoc.stanford.edu/}, \url{ftp://gong2.nso.edu/HA/haf/}, \url{https://cdaw.gsfc.nasa.gov/stereo/} and \url{http://sd-www.jhuapl.edu/secchi/wavelets/}.
\end{dat}

\begin{conflict}
The authors declare that they have no conflicts of interest.
\end{conflict}


\bibliographystyle{spr-mp-sola}
\bibliography{references}  

\end{article} 
\end{document}